\documentclass[aps,floatfix,showpacs,twocolumn, preprintnumbers,numbersect]{revtex4-1}
\usepackage[utf8]{inputenc}

\usepackage{bm}
\usepackage{color}
\usepackage{graphicx,epsfig}
\usepackage{amsfonts,amssymb,amsmath}
\usepackage{float}
\usepackage{xcolor}
\usepackage{multirow}
\usepackage{subcaption}
\usepackage{hyperref}
\usepackage{soul}
\usepackage{cancel}

\hypersetup{colorlinks,urlcolor=blue,allcolors=blue}

\usepackage{soul}

\usepackage{dcolumn}

\newcommand{\nc}{\newcommand}

\newcommand{\beq}{\begin{equation}}
\newcommand{\eeq}{\end{equation}}
\nc{\bfx}{{\bf x}}
\nc{\bfy}{{\bf y}}
\nc{\bfz}{{\bf z}}
\nc{\bfxh}{{\bf \hat{x}}}
\nc{\bfyh}{{\bf \hat{y}}}
\nc{\bfzh}{{\bf \hat{z}}}
\nc{\bfj}{{\bf j}}
\nc{\bfr}{{\bf r}}
\nc{\bfR}{{\bf R}}
\nc{\bfk}{{\bf k}}
\nc{\bfq}{{\bf q}}
\nc{\bfp}{{\bf p}}
\nc{\bfv}{{\bf v}}
\nc{\bfs}{{\bf s}}
\nc{\bfA}{{\bf A}}
\nc{\bfJ}{{\bf J}}
\nc{\bfsg}{{\bm \sigma}}
\nc{\bfvh}{{\bf \hat{v}}}
\nc{\bfqh}{{\bf \hat{q}}}
\nc{\low}{\delta_{\rm Low}}

\newcommand{\ket}[1]{| #1 \rangle}
\newcommand{\bra}[1]{\langle #1 |}

\nc{\swap}{\rightleftharpoons}
\nc{\grad}{{\bm \nabla}}

\newcommand{\bs}[1]{\ensuremath{\boldsymbol{#1}}}

\newcommand{\be}{\begin{equation}}
\newcommand{\ee}{\end{equation}}

\begin{document}

\title{Relativistic corrections for lepton-nucleus scattering in the short-time approximation}
\author{L.\ Andreoli$^{1,2}$}
\email{landreol@odu.edu}
\author{R.\ Weiss$^{3}$}
\email{ronen@wustl.edu}
\author{G. \ Chambers-Wall$^{3}$}
\email{chambers-wall@wustl.edu}
\author{A. \ Gnech$^{1,2}$}
\email{agnech@odu.edu}
\author{S.\ Pastore$^{3,4}$}
\email{saori@wustl.edu}
\author{M.\ Piarulli$^{3,4}$}
\email{m.piarulli@wustl.edu}
\author{S.\ Gandolfi$^5$}

\affiliation{
$^1$\mbox{Department of Physics, Old Dominion University, Norfolk, VA 23529, USA}\\
$^2$\mbox{Theory Center, Jefferson Lab, Newport News, VA 23610, USA}\\
$^3$\mbox{Department of Physics, Washington University in Saint Louis, Saint Louis, MO 63130, USA}\\
$^4$\mbox{McDonnell Center for the Space Sciences at Washington University in St. Louis, MO 63130, USA}\\
$^5$\mbox{Theoretical Division, Los Alamos National Laboratory, Los Alamos, NM 87545, USA}\\
}

\begin{abstract}
We present an approach for including relativistic corrections in lepton-nucleus scattering calculations within the Short-Time Approximation (STA). Previous \emph{ab-initio} studies employed electromagnetic currents expanded in powers of $q/m$, where $q$ is the momentum transfer and $m$ is the nucleon mass, restricting their validity to low-$q$ kinematics. We adopt an expansion scheme that treats the initial nucleon momentum perturbatively while allowing for arbitrary momentum transfer, thereby extending the applicability of the STA to high-$q$ regimes. Additionally, we incorporate a relativistic treatment of the two-nucleon final-state energies. Calculations for $^3$He and $^4$He inclusive electron scattering cross sections show a substantial improvement over previous results, achieving good agreement with experimental data in the quasi-elastic region for both low- and high-momentum transfer.
\end{abstract}

\maketitle

\section{Introduction}
\label{sec:intro}
In recent years, significant progress has been made in calculating inclusive cross sections for lepton scattering from nuclei~\cite{Rocco:2020jlx,Bacca:2014tla} using the microscopic, or {\it ab initio}, approach~\cite{10.3389/fphy.2020.00379}. This framework allows for the inclusion of many-nucleon correlations and electroweak currents, which are essential for delivering theoretical predictions that explain the available experimental data. Specifically, nuclear wave functions are determined by solving the many-nucleon Hamiltonian, which typically includes a kinetic term arising from the sum of the kinetic energies of individual nucleons, as well as two- and three-body potentials that describe the correlations between nucleons in pairs and triplets. The interaction with external electroweak probes, such as electrons and neutrinos, is expressed by one- and two-nucleon current operators, which describe the coupling of the probe with individual nucleons and pairs of correlated nucleons.

Due to the increasing computational cost associated with a larger number of nucleons, the {\it ab initio} method has primarily been applied to light nuclei~\cite{Carlson:1997qn,Bacca:2014tla,Carlson:2001mp,Carlson:1992ga,Lovato:2015oea,Lovato:2016gkq,Lovato:2017cux,Lovato:2020kba,Efros:1994iq,Bacca:2009vp,Orlandini:2013eya}. Only recently have calculations for systems with a mass number $A=40$ appeared in the literature~\cite{Sobczyk:2023sxh,Sobczyk:2021dwm}. These studies have highlighted the importance of many-nucleon effects in accurately explaining the available experimental data.

Despite these advancements, there are two key limitations within this approach. First, traditional methods are generally limited to studying inclusive processes. Specifically, these methods rely on integral properties of the nuclear electroweak response functions, which are extracted by unfolding the integral transforms. For example, QMC calculations, specifically Green's Function Monte Carlo (GFMC) calculations, exploit the Euclidean transform~\cite{Carlson:2001mp,Carlson:1992ga,Lovato:2015oea,Lovato:2016gkq,Lovato:2017cux,Lovato:2020kba}, while the coupled-cluster (CC) approach uses the Lorentz integral transform (LIT) method~\cite{Efros:1994iq,Bacca:2009vp,Orlandini:2013eya}. 

Second, the {\it ab initio} framework is inherently nonrelativistic, as it solves the many-body Schr\"odinger equation. Relativistic effects in the electroweak currents are typically treated perturbatively by expanding the covariant single-nucleon currents in powers of $q/m$, where $q$ is the momentum transferred by the external electroweak probe and $m$ is the nucleon mass. Relativistic effects may also be phenomenologically included in two-nucleon correlations. In fact, the two-nucleon interactions typically adopted in QMC calculations, {\it i.e.}, the Argonne $v_{18}$~\cite{Wiringa:1994wb}, contain parameters that are fitted to nucleon-nucleon scattering data, including data that involve center-of-mass energy up to around $300$ MeV, and reproduce the phase shifts up to $\sim 1 $ GeV where relativistic effects may become significant. 

Despite the {\it approximated} inclusions of relativistic effects, this approach fails to agree with the data at high values of momentum transfer. The importance of a proper relativistic treatment of the electroweak currents has recently been highlighted in the context of electron scattering from trinucleon systems~\cite{Andreoli:2021cxo}, where QMC-based calculations -- namely the GFMC and the Short-Time-Approximation (STA) -- of cross sections were compared with those obtained with the Spectral Function method~\cite{Benhar:1994hw,Rocco:2015cil,Rocco:2018mwt}. The latter adopts fully covariant single nucleon currents and accounts for relativistic kinematics in the final-state energies, thus correctly reproducing the experimental data at high $q$ values, while QMC calculations were in agreement with data up to values of $q\simeq 600$ MeV$/c$. 

To address these limitations, two approaches have been explored for incorporating relativistic effects in nonrelativistic {\it ab initio} QMC calculations. One relies on boost transformations~\cite{PhysRevC.72.011002}. This approach has been applied, for example, in combination with nuclear electroweak responses obtained from GFMC calculations, leading to improvements in the description of scattering at higher momentum transfers~\cite{Rocco:2018tes,Nikolakopoulos:2023zse}. 

The second approach, which is the focus of this work, is based on the Short-Time Approximation (STA)~\cite{Pastore:2019urn} factorization scheme, where the external probe interacts only with pairs of correlated nucleons. This enables us to extend the {\it ab initio} approach for nuclear electroweak responses and cross sections to larger nuclei without losing essential two-nucleon dynamics. The STA has been validated through comparisons with exact theoretical results and experimental data of electron scattering from light nuclei with mass numbers $A=3,\, 4$ and $12$~\cite{Pastore:2019urn,Andreoli:2021cxo,Andreoli:2024ovl,King:2024zbv}, showing good agreement with the data up to values of momentum transfer $q\sim 600$ MeV$/c$. The STA factorization scheme consistently incorporates two-nucleon correlations and currents while isolating the interaction vertex, thus allowing for a direct treatment of relativistic effects. 
 
Specifically, relativistic effects are included in the single-nucleon current operator and in the treatment of the two-nucleon final state kinematics. The nonrelativistic reduction of the covariant single-nucleon electromagnetic current is carried out using an expansion scheme where the momentum transfer $q$ is no longer considered as a small perturbative scale, as in traditional calculations~\cite{Carlson:1997qn,Pastore:2008ui,Pastore:2009is,Pastore:2011ip}. Instead, it assumes that the momentum $q$ carried by the probe can be large and transferred to the scattered nucleon, while only the momentum of the struck nucleon inside the nucleus is small and is treated perturbatively. This  approach allows for a more accurate theoretical description of relativistic effects ensuring better agreement with experimental data at high $q$-values. We note that a similar approach has been developed, {\it{e.g.}}, in Ref.~\cite{Jeschonnek:1997dm} to study single nucleon electron scattering and $^2$H$(e,e^\prime p)n$ reactions. We will discuss some of the differences in the relevant section.

The theoretical framework developed in this work is directly relevant for current and future experimental programs in neutrino-nucleus  scattering~\cite{DUNE:2016hlj,DUNE:2022aul} and can be extended beyond the light nuclei studied here. Electron scattering provides a valuable testing ground for nuclear models applicable to neutrino physics, due to the wealth of  experimental data and the absence of complications related to leptonic energy reconstruction~\cite{Ankowski:2014}. Although the STA has been validated for nuclei up to $A=12$ within the Variational Monte Carlo (VMC) approach, its algorithm can be exported to computational methods utilized to study heavier systems. 
For the Deep Underground Neutrino Experiment (DUNE), which uses a $^{40}$Ar target, future developments of the STA approach in combination with the Auxiliary Field Diffusion Monte Carlo~\cite{Carlson:2014vla} many-body computational method, will allow for lepton-nucleus studies toward medium-mass nuclei contributing to a better modeling of neutrino-nucleus interactions at the few-GeV energy scale.
Continued development of theoretical approaches that properly account for relativistic effects and two-nucleon currents and correlations in nuclear systems beyond the light mass region are required to reduce uncertainties in neutrino energy reconstruction that are crucial for DUNE's oscillation measurements.

Similarly, this framework allows us to address observables  measured in ongoing and future electron scattering experiments at Jefferson Lab~\cite{Arrington:2021alx,Accardi:2023chb}.
Extension of the STA framework to compute the cross section measured in short-range correlation experiments~\cite{Arrington:2022sov} as well as Primakoff production of pseudoscalar mesons~\cite{PrimEx:2010fvg} are also possible and will be considered in future works. The development of the {\it ab-initio} technique such as the STA is crucial to understanding and describing the experiments with light nuclei at Jefferson JLab and in the future at EIC~\cite{AbdulKhalek:2021gbh}.

The inclusion of relativistic effects at the vertex within the framework of the STA is the focus of this paper. The paper is organized as follows. Section~\ref{sec:xsec} describes nuclear response functions and inclusive cross sections. In Section~\ref{sec:currents}, we discuss the covariant single-nucleon electromagnetic current and its nonrelativistic reduction suitable for large $q$-values. Section~\ref{sec:STA} outlines the STA method. The relativistic treatment of two-nucleon final state kinematics is discussed in Section~\ref{sec:relativistic_kinematics}. Results and comparisons with experimental data for inclusive electron scattering are provided in Section~\ref{sec:results}, while in Section~\ref{sec:conclusion} we summarize our conclusions and outline the next stage in this program of extending the kinematic reach of the STA.

\section{Electron-nucleus scattering cross section}
\label{sec:xsec}

The inclusive double-differential cross section for electron scattering from a nucleus is  written in terms of the longitudinal, $R_L$, and transverse, $R_T$, response functions as:
\vspace{-6pt}
\begin{align}
\label{eq:diff_cross_section}
    \begin{split}
       \left(\frac{d^2\sigma}{d\varepsilon^\prime d\Omega}\right)_e=&\left(\frac{d\sigma}{d\Omega}\right)_M\left[\left(\frac{Q^2}{{q}^2}\right)^2R_L(q,\omega)\right.
        \\
        &\left.+\left(\tan^2{\frac{\theta}{2}}+\frac{1}{2}\frac{Q^2}{{q}^2}\right) R_T({q},\omega) \right],
    \end{split}
\end{align}
where $\varepsilon'$ is the electron's final energy and $\Omega$ is the scattering solid angle. The Mott cross section is given by
\begin{equation}
    \left(\frac{d\sigma}{d\Omega}\right)_M = \left[\frac{\alpha \cos \left(\theta / 2\right)}{2 \varepsilon^{\prime} \sin ^{2}\left(\theta / 2\right)}\right]^{2} \, ,
\end{equation}
where $\boldsymbol{q}=\boldsymbol{p}_e-\boldsymbol{p}^\prime_e$ and $q$ denote the three-momentum transfer and its magnitude, respectively, and $\boldsymbol{p}_e$ ($\boldsymbol{p}^\prime_e$) is the momentum of the incoming (outgoing) electron.  The energy transfer is denoted by $\omega=\varepsilon -\varepsilon^\prime$, while $Q^2=q^2-\omega^2$, and $\theta$ is the electron scattering angle.
The longitudinal and transverse response functions of Eq.~(\ref{eq:diff_cross_section}) describe the transition from an initial, $\ket{\Psi_0}$, to a final nuclear state, $\ket{\Psi_f}$, of a nucleus with mass number $A$ and initial (final) energy $E_0$ ($E_f$).
Specifically, 
\beq \label{eq:R_L}
R_L(q,\omega) = R^{00}(q,\omega) \, , 
\eeq
and
\beq \label{eq:R_T}
R_T(q,\omega) = R^{xx}(q,\omega)+R^{yy}(q,\omega) \, , 
\eeq
where 
\begin{align}\label{eq:hardonictensor}
    \begin{split}
        R^{\mu\nu}(\boldsymbol{q},\omega)=&\sum_f\bra{\Psi_0}J^{\mu\dagger}(\boldsymbol{q},\omega)\ket{\Psi_f}\bra{\Psi_f}J^\nu(\boldsymbol{q},\omega)\ket{\Psi_0}
        \\
        &\times\delta(E_0+\omega-E_f)\, ,
    \end{split}
\end{align}
and the sum runs on all possible final hadronic states, both bound and in the continuum.

The current $J^\mu(\boldsymbol{q},\omega)$ describes the interaction between the probe and the nucleus and it is expanded as a sum of many-nucleon operators as
\beq
    J^\mu=\sum_i j^\mu_i+\sum_{i<j} j^\mu_{ij}+\ldots \, , 
\eeq
where $j^\mu_i$ is a one-body operator acting on the $i$-th nucleon and the sum runs over all the nucleons in the system. The two-body component, $j^\mu_{ij}$, acts on nucleon correlated pairs, $ij$, and the corresponding sum goes over all pairs in the system. As in conventional calculations, we neglect terms that act on more than two nucleons and choose $\bs{q}$ to be along the $z$-axis. 

The focus of this work is on an improved treatment of the one-body current operators at high values of momentum transfer $q$. For two-body currents, we employ the most recent formulation described at length in Refs.~\cite{Carlson:2014vla,Bacca:2014tla,Marcucci:2005zc,Pastore:2019urn,Andreoli:2024ovl} and references therein. These will not be discussed further here; interested readers are referred to those sources for more details. 

The computational methods used in this work are described in Refs.~\cite{Pastore:2019urn,Andreoli:2024ovl}. Our calculations employ a nuclear Hamiltonian consisting of the Argonne $v_{18}$ two-nucleon interaction \cite{Wiringa:1994wb} together with the Urbana IX three-nucleon interaction \cite{Pudliner:1995wk}. The ground state wave function includes two- and three-body correlations induced by the Hamiltonian.

\section{One-Body Electromagnetic Current}
\label{sec:currents}
The covariant relativistic expression for the one-body electromagnetic current operator is given by~\cite{Carlson:1997qn}
\begin{align}\label{eq:relvector}
    j^\mu=\bar{u}(\boldsymbol{p}^\prime s')\left[{F}_1\gamma^\mu+i\sigma^{\mu\nu}q_\nu\frac{{F}_2}{2m}\right]u(\boldsymbol{p}s),
\end{align}
where $\boldsymbol{p}$ ($\boldsymbol{p}^\prime$) is the initial (final) nucleon momentum with mass $m$ and spin $s$ ($s'$) and we omit, here and in the following, the index that specifies the $i$-th nucleon. The Dirac spinors and gamma matrices are denoted by $u$ and $\gamma^\mu$, respectively, with $\bar{u}=u^\dagger \gamma^0$ and $\sigma^{\mu\nu}=\frac{i}{2}[\gamma^\mu,\gamma^\nu]$. The $4$-vector energy-momentum transfer is given by $q^\nu=(\omega,\bs{q})$. 

The Dirac and Pauli form factors~\cite{Hyde-Wright:2004}, ${F}_1(Q^2)$ and $F_2(Q^2)$,  characterize the electromagnetic structure of the nucleon and are expressed in terms of the electric and magnetic Sachs form factors $G_{E/M}^{p,n}(Q^2)$~\cite{Ernst:1960} of the proton and neutron,
\begin{align}
\label{eq:sachs}
    F_1^{p,n}=\frac{G_E^{p,n}+\tau G_M^{p,n}}{1+\tau},\quad F_2^{p,n}=\frac{G_M^{p,n}-G_E^{p,n}}{1+\tau},
\end{align}
where $\tau=Q^2/4m^2$ is a kinematic factor. The nucleon's identity -- whether it is a proton ($p$) or a neutron ($n$) -- is specified by the nucleon isopsin projection operator $(1\pm\tau_z)/2$, with $\tau_z$ denoting the third component of the nucleon isospin operator. In what follows, for ease of reading, we omit the $p/n$ superscript and implicitly assume the appropriate isospin projection operator acting on the nucleon. 

\subsection{Low-momentum transfer nonrelativistic reduction}

Traditional many-body calculations~\cite{Carlson:1997qn,Bacca:2014tla} adopt a nonrelativistic reduction of the single-nucleon covariant current given in Eq.~(\ref{eq:relvector}). The expansion is based on the assumption that both $p$ and $p'$ (and thus $q$) are small compared to the nucleon mass. At leading order (LO), this approach leads to the following expressions in coordinate space for the charge, $\rho=j^0$, and current, $\boldsymbol{j}_\perp$, operators:
\begin{align} \label{eq:charge_NR_LO}
    \rho=&{G_E} e^{i\boldsymbol{q}\cdot\boldsymbol{r}} \, ,
\end{align}
\begin{align} \label{eq:transverse_NR_LO}
\boldsymbol{j}_\perp=&e^{i\boldsymbol{q}\cdot\boldsymbol{r}}\left[\frac{G_E}{m}\boldsymbol{p}^\perp-i\frac{G_M}{2m}(\boldsymbol{q}\times\boldsymbol{\sigma})\right] \, ,
\end{align}
where $\bs{O}^\perp = \bs{O}-(\bs{O}\cdot\hat{\bs{q}}) \hat{\bs{q}}$ is the component of the operator $\bs{O}$ perpendicular to the momentum transfer $\bs{q}$, and $\boldsymbol{r}$ is its spatial coordinate.
Additionally, in coordinate space $\bs{p} = -i \nabla$ and acts directly on the wave function (for sub-leading terms involving powers of $\bs{p}$, particular care must be taken when performing the Fourier transform from momentum to coordinate space).

Typically, in addition to the LO term given above, the single-nucleon charge operator includes the Darwin-Foldy and spin-orbit terms, also accounting for relativistic effects~\cite{Carlson:1997qn}:
\begin{align} 
\label{eq:charge_DF}
    \rho=&\left[\frac{G_E}{\sqrt{1+Q^2/4m^2}}
    -i\frac{2G_M-G_E}{4m^2}\boldsymbol{q}\cdot(\boldsymbol{\sigma}\times\boldsymbol{p})\right] e^{i\boldsymbol{q}\cdot\boldsymbol{r}}   \, .
\end{align}

Using the expansion parameter discussed above, that is $q/m \sim p/m \sim p'/m \ll 1$, subleading terms of the one-body charge and current operators, beyond LO, have been derived, {\it e.g.}, in Refs.~\cite{Pastore:2009is,Pastore:2011ip}. However, the expansion is limited to low-momentum-transfer kinematics and consequently fails to accurately describe processes at high energy. An appropriate expansion scheme for these regimes is described in the next subsection.

\subsection{High-momentum transfer nonrelativistic reduction}
\label{sec:relativistic_current}
For momentum transfer $q \sim m$ or greater, relativistic effects become important in two main ways. First, the electromagnetic current operators require a proper nonrelativistic reduction.  Second, the final states, which play a significant role in the nuclear response functions, may include high-momentum knocked-out nucleons, whose kinetic energies should be calculated relativistically for accurate results.
In this section, we focus on determining the nonrelativistic reduction of the electromagnetic currents at high-momentum transfer. The results obtained here are general and applicable in any microscopic modeling of the nucleus.  The treatment of final states' relativistic kinematics within the STA framework is discussed separately in Sec.~\ref{sec:relativistic_kinematics}.

Our starting point is the covariant expression of the one-body current operator given in Eq.~(\ref{eq:relvector}). At low-momentum transfer $q$, its nonrelativistic reduction at leading order is given in Eqs.~\eqref{eq:charge_NR_LO} and~\eqref{eq:transverse_NR_LO} which are accurate for $q\ll m$. Sub-leading terms extend the currents' validity up to moderate $q$, but the expansion breaks down at larger momenta since it assumes $q/m\ll1$. Even the inclusion of sub-leading terms fails to overcome this limitation. To address this, we adopt a different strategy -- introduced in Ref.~\cite{Jeschonnek:1997dm} -- in which only the initial nucleon momentum $p$ is assumed to be small. 
This is justified in the quasi-elastic kinematic regime where the struck nucleon is initially at low momentum but receives a momentum $q$ from the probing lepton, which can be large. Thus, we avoid the assumption that the nucleon final momentum $p^\prime$ is small and expand the covariant current only in $p/m$. The resulting expressions are then valid for any value of momentum transfer $q$ provided that $p/m\ll1$. 

To carry out this nonrelativistic reduction we begin by defining the spinors and their normalizations
\beq \label{eq:spinor}
u(\bs{p}s) = N(p) \begin{pmatrix}
           1 \\
           \frac{\bs{\sigma}\cdot \bs{p}}{E+m} 
         \end{pmatrix}
\chi_s \, ,
\eeq
where $\chi_s$ and $E=\sqrt{p^2+m^2}$ are the nucleon spin state and energy, respectively,  while $N$ is a normalization factor chosen so that $u^\dagger u = 1$. Specifically, for a nucleon with momentum $p$ and energy $E$, the normalization factor reads:
\be \label{eq:N_p}
N(p) = \sqrt{\frac{E+m}{2E}}  \, .
\ee
An expansion of the normalization factor in $p/m$ yields $N(p) \rightarrow 1$ at leading order. 

For the final state, where $\bs{p'}=\bs{p}+\bs{q}$, we make no assumption on the size of $p^\prime$ and $q$. Thus, at leading order the normalization factor of the final nucleon spinor reads: 
\beq \label{eq:N_p'}
N(p^\prime)=\sqrt{\frac{E^\prime+m}{2E^\prime}} \rightarrow  \sqrt{\frac{\sqrt{q^2+m^2}+m}{2\sqrt{q^2+m^2}}} \equiv \alpha(q) \, .
\eeq
Sub-leading corrections arise from further expanding  in $p/m$. 

This expansion scheme is also applied to the factors $(E+m)^{-1}$ and $(E'+m)^{-1}$ entering the spinors, and to the energy transfer $\omega$ defined as:
\be \label{eq:omega}
\omega=E'-E = \sqrt{p'^2+m^2}-\sqrt{p^2+m^2} \, .
\ee
We note that this formulation assumes that both the initial and final nucleons are on-shell. While different conventions exist, the impact of off-shell corrections near the quasi-elastic peak is expected to be minimal~\cite{DeForest:1983}.

The nucleon form factors ${F}_1(Q^2)$ and $F_2(Q^2)$ appearing in the covariant expression of the electromagnetic current -- Eq.~(\ref{eq:relvector}) -- must also be treated consistently since they depend on the nucleon momentum through $Q^2$. At leading order in $p/m$, this effectively corresponds to evaluating the form factors at the quasi-elastic peak, that is at $p=0$, or equivalently at $Q^2=Q_{qe}^2$ with
\be
Q^2_{qe}=2m\left(\sqrt{q^2+m^2}-m\right) \, .
\ee
Then the form factor expressions become: 
\be
G_E(Q^2_{qe}) = F_1(Q^2_{qe})-\tau_{qe} F_2(Q^2_{qe}) \, ,
\ee
\be
G_M(Q^2_{qe})=F_1(Q^2_{qe})+F_2(Q^2_{qe}) \, ,
\ee
with
\be
\tau_{qe}= \frac{Q^2_{qe}}{4m^2} =
\frac{\sqrt{q^2+m^2}-m}{2m}  \, .
\ee

By expanding the covariant electromagnetic current in $p/m$ while treating the momentum transfer $q$ exactly, we obtain the following expressions at leading order (LO), $\sim (p/m)^0$: 
\begin{eqnarray}
\label{eq:rho_p0}
\rho^{(0)} &=& \alpha(q) G_E(Q^2_{qe})e^{i\boldsymbol{q}\cdot\boldsymbol{r}}\, ,\\
\label{eq:current_p0}
\bs{j}_\perp^{(0)}  &=&
\frac{2m\tau_{qe}}{q^2} G_M(Q^2_{qe}) \alpha(q)   i  (\bs{\sigma} \times \bs{q}) e^{i\boldsymbol{q}\cdot\boldsymbol{r}}\, .
\end{eqnarray}

The next-to-leading-order (NLO) expressions include terms of the order $\sim(p/m)^1$ and are given by:
\begin{widetext}
\begin{eqnarray}
\label{eq:rho_p1}
\rho^{(1)} & = & \frac{\alpha(q) \tau_{qe}}{m^2(2\tau_{qe}+1)^2(\tau_{qe}+1)} \left[
(2\tau_{qe}+1) G_M   - \tau_{qe}G_E  \right] \bs{q}\cdot \bs{p}
-
\frac{\alpha(q)}{(\tau_{qe}+1)} \frac{2G_M - G_E}{4m^2} 
i\bs{q}\cdot(\bs{\sigma}\times\bs{p})\, ,     \\
\bs{j}^{(1)}_\perp &= &
 \left\{ 
 \frac{\alpha(q)}{m(\tau_{qe}+1)} \left[  G_E
+ \frac{\tau_{qe}  }{2} G_M \right]
\right\} \bs{p}^\perp
+
 \left\{ 
  \frac{\alpha(q) \tau_{qe}}{m(\tau_{qe}+1)} 
  \left[ \frac{1}{2}G_E - G_M \right]
\right\}   i(\bs{\sigma} \times \bs{p})^\perp
 \nonumber \\ &+&
\left\{ 
\frac{\alpha(q)}{4m^3(\tau_{qe}+1)^2 (2\tau_{qe}+1)^2} \left[
 - G_E \frac{1}{2}(2\tau_{qe} +1)^2 
+G_M (2\tau_{qe}^2-1)
\right]
\right\} \left(\bs{p}\cdot\bs{q}\right)(i\bs{\sigma}\times\bs{q}) 
\nonumber \\ &+&
\left\{ 
 \frac{\alpha(q)}{8m^3(\tau_{qe}+1)^2}
(G_M-G_E)
\right\} ( \bs{\sigma}\cdot\bs{q})(i\bs{q}\times\bs{p}) \, ,
\label{eq:current_p1}
\end{eqnarray}
\end{widetext}
where we dropped the overall $e^{i\boldsymbol{q}\cdot\boldsymbol{r}}$ factor for ease of reading. Note that also at NLO the Sachs form factors are evaluated at the quasielastic peak, so in the equations above $G_{E/M}\equiv G_{E/M}(Q_{qe}^2)$. Corrections beyond this approximation will appear at the next order. 

 The nucleonic form factors $G_{E/M}$ are described by empirical functions, typically constrained through fits to experimental data. In principle, their expansion around the quasielastic peak introduces terms that should be included for consistency.  For example, terms linear in $p$, when  combined with the operator structures of Eqs.~\eqref{eq:rho_p0} and \eqref{eq:current_p0}, would contribute at NLO. While these contributions could be determined explicitly for a given parameterization, this lies beyond the scope of this work.

The expansion presented here closely follows the approach developed in Ref.~\cite{Jeschonnek:1997dm}, but for some differences arising from the use of different conventions. In particular, we adopt the normalization condition $u^\dagger u =1$, which leads to the normalization factor given in Eq.~\eqref{eq:N_p}. The same convention is effectively used, for example, in Ref.~\cite{Krebs:2019aka} (see Eqs. (2.2), (2.3), and (2.5) therein). This convention is consistent with the definition of cross section given in Eq.~\eqref{eq:diff_cross_section}. Ref.~\cite{Jeschonnek:1997dm}, on the other hand, uses the normalization factor $N=\sqrt{(E+m)/(2m)}$, a convention also adopted in Ref.~\cite{Amaro:1995kk} (see Eq.~(A.6) in that reference). As a result, their corresponding expression for the cross section includes additional kinematics factors compared with the definition given in Eq.~\eqref{eq:diff_cross_section}.

While the different conventions lead to distinct expressions in the response functions and currents, they do not affect the cross sections at LO in $p/m$. For example, Ref.~\cite{Jeschonnek:1997dm} reports an increase in the longitudinal response function (relative to the longitudinal response calculated with the LO charge operator of Eq.~\eqref{eq:charge_NR_LO} obtained by expanding in $q/m$), whereas we find a decrease. Despite this, the corresponding cross sections agree at this order. Another difference that affects the formal expressions at NLO and beyond is related to the treatment of the  energy transfer $\omega$ given in Eq.~\eqref{eq:omega}, which we expand to allow for a more efficient implementation in QMC calculations. 

As a consistency check, we have verified that our conventions for both the spinor normalization and the cross-section  reproduce the well-known Rosenbluth cross section in the limit of a single  nucleon at rest in the initial state. This verification is detailed in App.~\ref{sec:ros_cross_section}

Here, we have focused exclusively on the one-body current operator. A similar treatment should be in principle applied to two-body currents. However, within the low-momentum transfer expansion scheme, the two-body vector and charge currents scale as $p/m$~\cite{Pastore:2011ip,Krebs:2019aka}, therefore, corrections to their standard formulations are expected to be small overall. For an expansion of two-body meson exchange currents see, {\it e.g.}, Ref.~\cite{Amaro:1998ta}.

\subsection{Comparing low- and high-momentum transfer nonrelativistic reductions}
\label{sec:comparing}

In this section, we compare the low-momentum transfer nonrelativistic reduction at LO of the single nucleon charge and current operators -- given in Eqs.~\eqref{eq:charge_NR_LO} and~\eqref{eq:transverse_NR_LO} -- with the corresponding expressions obtained in the high-momentum transfer regime -- Eqs.~\eqref{eq:rho_p0} and~\eqref{eq:current_p0}. 

\begin{figure} \begin{center} 
\includegraphics[width=\linewidth]{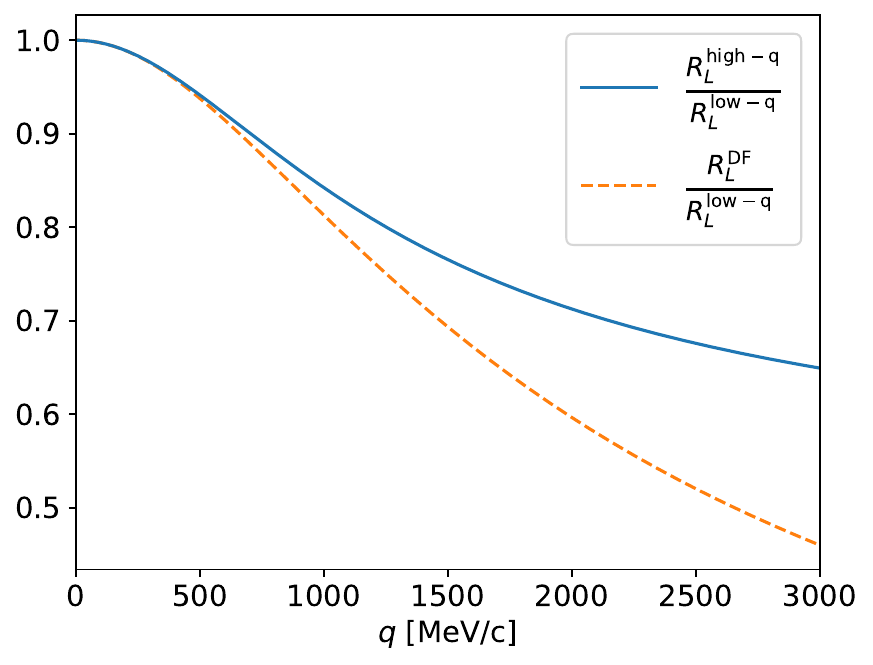}
\caption{\label{fig:R_L_factors} 
Ratio of the longitudinal responses obtained using: {\it i}) the high- and low-$q$ expansion schemes for the one-body charge orator at LO (blue solid line); and {\it ii}) the Darwin-Foldy correction to the charge operator and the low-$q$ expansion at LO (orange dashed line).   
See text for explanation.}
\end{center}\end{figure}

We start with the charge operator. Here, the main difference is the additional factor $\alpha(q)$ in the LO high-$q$ result of Eq.~\eqref{eq:rho_p0}. Since $\alpha(q) \rightarrow 1$ for $q\rightarrow 0$, the two results coincide for $q\ll m$. However, $\alpha(q)<1$ for all values of $q$, and its deviation from unity increases with increasing $q$, leading to a suppression in the longitudinal response at LO. Specifically, the longitudinal response -- which involves a current-current correlator -- scales as $\alpha(q)^2$ in the high-$q$ expansion scheme, whereas in the low-$q$ reduction the scaling prefactor is simply one. The ratio of the longitudinal responses obtained in the two different schemes, namely
\be 
\frac{R_L^{\rm high-q}}{R_L^{\rm low-q}}\propto \alpha(q)^2 \, ,
\ee
is plotted in Fig.~\ref{fig:R_L_factors} as a function of $q$. The reduction becomes increasingly pronounced at high $q$. For example, $\alpha^2=0.9$ for $q=700$ MeV$/c$, and $\alpha^2=0.77$ for $q=1500$ MeV/$c$, and, in the limit of $q\gg m$, $\alpha^2\rightarrow 1/2$.

We can also compare the low-$q$ result at LO with the Darwin-Foldy contribution to the charge operator, given by the first term in Eq.~\eqref{eq:charge_DF}. In this case, the resulting ratio of longitudinal responses is proportional to 
\be 
\frac{R_L^{\rm DF}}{R_L^{\rm low-q}}\propto \frac{1}{(1+Q_{qe}^2/4m^2)}\, ,
\ee
and it is also plotted in Fig.~\ref{fig:R_L_factors} for comparison. The LO high-$q$ expression (solid blue line) and the Darwin-Foldy (orange dashed line) term agree well up to moderate values of momentum transfer. Differences arise at $q\gtrsim 500$ MeV/$c$. In this regime, the Darwin-Foldy prefactor decreases more rapidly than $\alpha(q)$, resulting in a stronger suppression in the longitudinal response. Consequently, the high-$q$ nonrelativistic reduction leads to comparatively larger longitudinal response functions. For example, at $q=1500$ MeV$/c$ there is a $\sim10\%$ enhancement. Moreover, in the limit of $q\gg m$, the Darwin-Foldy term goes to zero, unlike $\alpha^2(q)$, leading to significant differences. 

We now turn our attention to the transverse current operator and compare its LO low-momentum transfer expression, Eq.~\eqref{eq:transverse_NR_LO}, with the LO high-momentum result, Eq.~\eqref{eq:current_p0}. In the limit of $q\rightarrow 0$, the factor $(2m\tau_{qe})/q^2\rightarrow 1/2m$, and the expression in Eq.~\eqref{eq:current_p0} reduces to the second term in Eq.~\eqref{eq:transverse_NR_LO}. The first term in Eq.~\eqref{eq:transverse_NR_LO}, proportional to $\boldsymbol{p}^\perp$, appears at next-to-leading-order in the high-$q$ reduction and is expected to contribute less at large momentum transfer. Its coefficient matches the corresponding term in Eq.~\eqref{eq:current_p1} in the $q\rightarrow 0$ limit.

\begin{figure} \begin{center} 
\includegraphics[width=\linewidth]{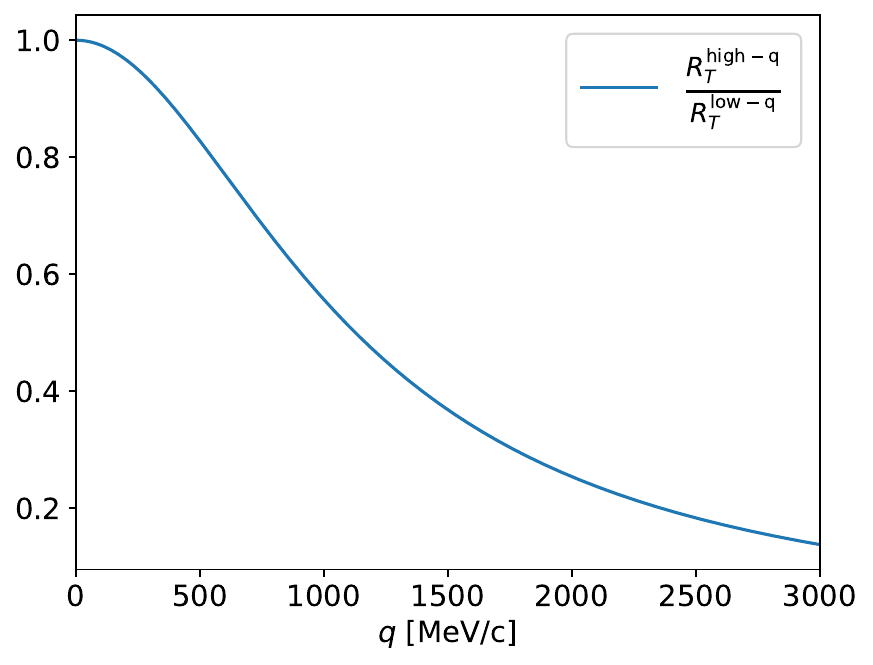}
\caption{ \label{fig:R_T_factors} Ratio of the transverse responses obtained using the LO high-$q$ and LO low-$q$ expansion schemes for the one-body current operator. See text for explanations.}
\end{center}\end{figure}

Focusing on the dominant contributions at large $q$, the prefactor $(2m\tau_{qe})\alpha(q)/q^2$ in the LO high-$q$ expansion should be compared to the factor $1/2m$ appearing in Eq.~\eqref{eq:transverse_NR_LO}. Therefore, the relative suppression factor in  the LO transverse response function is 
\be 
\frac{R_T^{\rm high-q}}{R_T^{\rm low-q}}\propto\left( \frac{4m^2\tau_{qe} \alpha(q)}{q^2} \right)^2. 
\ee
This factor, plotted  in Fig.~\ref{fig:R_T_factors}, is always less than one and decreases with increasing values of $q$, leading to a suppression of the transverse response. Already at $q\approx 400$ MeV/$c$ the suppression is around $10\%$. At $q=700$ MeV/$c$, the factor is $\sim 0.71$,  and it drops to $\sim0.37$ at  $q=1500$ MeV$/c$.

In the limit of $q\gg m$, the factor scales as $2m^2/q^2$, vanishing for large $q$. This leads to a significant reduction in the transverse response function due to relativistic effects -- an effect that cannot be captured by the low-$q$ nonrelativistic reduction. 

\section{Short-time approximation}
\label{sec:STA}

In this section, we briefly sketch the Short-Time-Approximation (STA) algorithm in the context of Variational Monte Carlo calculations and defer the interested reader to Refs.~\cite{Pastore:2019urn,Andreoli:2024ovl} for more details. We also compare our results with those of Refs.~\cite{Pastore:2019urn,Andreoli:2021cxo}, which used the low-$q$ nonrelativistic reduction of the single-nucleon covariant current.

The STA evaluates the response functions within a factorization scheme that includes two-nucleon dynamics at the interaction vertex~\cite{Pastore:2019urn}. The response function, Eq. \eqref{eq:hardonictensor}, can be expressed as
\begin{align}
\begin{split}
    R^{\mu\nu}(\boldsymbol{q},\omega)=&\int_{-\infty}^\infty\frac{dt}{2\pi}e^{i(\omega+E_0)t}
    \\
    &\times\bra{\Psi_0}J^{\mu\dagger}(\boldsymbol{q},\omega)e^{-iHt}J^\nu(\boldsymbol{q},\omega)\ket{\Psi_0}\, ,
    \end{split}
\end{align}
where the sum over the final states and the energy-conserving delta are replaced by the real-time propagator. For short times, the propagator can be expanded to retain only correlations involving two nucleons that interact with the probe via one- and two-body currents. The corresponding current-current correlator is approximated as:

\begin{align} \label{eq:currcurrcorr}
    \begin{split}
        J^{\mu\dagger}e^{-iHt}J^\nu\approx&\sum_i j^{\mu\dagger}_ie^{-i\,h_{ij}t}j^\nu_i+\sum_{i\neq j}j^{\mu\dagger}_ie^{-i\,h_{ij}t}j^\nu_j
        \\
        &+\sum_{i\neq j}\left(j_i^{\mu\dagger}e^{-i\,h_{ij}t}j_{ij}^\nu+j^{\mu\dagger}_{ij}e^{-i\,h_{ij}t}j_i^\nu\right)
        \\
        &+\sum_{i<j} j_{ij}^{\mu\dagger}e^{-i\,h_{ij}t}j_{ij}^\nu \, , 
    \end{split}
\end{align}
where $h_{ij}=t_i+t_j+v_{ij}$ includes the single-nucleon kinetic energies and the two-body interaction $v_{ij}$, the Argonne $v_{18}$ in the present study.
The second line represents the so called \textit{interference term} between one- and two-body currents, which is known to provide important corrections to the transverse response function~\cite{Carlson:2001mp}.

Because the correlated $ij$ pair is factorized from the remaining $A-2$ nucleons, the response function can be written in terms of a response density $D(e,E_{\mathrm{c.m.}})$:
\begin{align}
\label{eq:integration_eE}
    \begin{split}
    R^{\mu\nu}=&\int de\int_0^\infty dE_{\mathrm{c.m.}}D^{\mu\nu}(e,E_{\mathrm{c.m.}})
    \\
    &\times\delta(\omega+E-e-E_{\mathrm{c.m.}}) \, .
    \end{split}
\end{align}
where $e$ is the relative energy of the active pair, $E_{\mathrm{c.m.}}$ its center-of-mass energy in the final state, and $E$ its average initial energy (set to zero in the present work). 

The corresponding response density involves the two-nucleon propagator $g$:
\begin{align}
\label{eq:density_eE}
D^{\mu\nu}&(e,E_{\mathrm{c.m.}}) =
4 m^{3/2} \sqrt{E_{\mathrm{c.m.}}} 
\int d^3R' d^3r' d^3R d^3r d^{3}R_{A-2}
\nonumber \\ & \times
\sum_{\alpha_{A-2}} 
\sum_{\alpha_1,\alpha_2}
\sum_{\alpha_1',\alpha_2'}
\Big[ \langle \Psi_i | j_L^{\mu\dagger} | \bs{R}', \bs{r}' ,\alpha_1' \alpha_2', \bs{R}_{A-2}, \alpha_{A-2} \rangle
\nonumber \\ &\times
g(e,E_{\mathrm{c.m.}},\bs{r},\bs{r}',\bs{R},\bs{R}',\alpha_1,\alpha_2,\alpha_1',\alpha_2')
\nonumber \\ &\times
\langle \bs{R}, \bs{r} ,\alpha_1 \alpha_2, \bs{R}_{A-2} , \alpha_{A-2} | 
 j_R^\nu | \Psi_i \rangle \Big]\, ,
\end{align}
where $\bs{r}$ and $\bs{r}'$ describe the relative coordinate of the active pair, $\bs{R}$ and $\bs{R}'$ its c.m. coordinate, and $\alpha_1$, $\alpha_1'$, $\alpha_2$ and $\alpha_2'$ the spin and isospin of the pair. $\bs{R}_{A-2}$ is for the location of the remaining particles and $\alpha_{A-2}$ for their spin and isospin.  The abbreviated notation with $j_L$ and $j_R$ is used to indicate the sum of all contributions from Eq.~\eqref{eq:currcurrcorr}.

In its full form, the propagator $g$ includes  continuum and bound two-nucleon states of the Argonne $v_{18}$, thus ensuring consistency between the two-nucleon correlations and two-nucleon electromagnetic currents at the interaction vertex, required for a consistent determination of the interference term~\cite{Pastore:2019urn}. 

In the present work, however, we employ a simplified free-propagator where final state interactions are neglected and the nucleons in the final state are described by plane-waves. This approximation allows us to focus on relativistic corrections at the one-body vertex  but it introduces an inconsistency: the interference term between one- and two-body currents is no longer treated consistently with two-nucleon correlations. The effects of this inconsistency and the improvement expected when using a correlated-propagator are discussed in Ref.~\cite{Pastore:2019urn} and will be revisited in future work. 

Using a free-propagator, the response density reduces to:  
\begin{align}
&D^{\mu\nu}(e,E_{\mathrm{c.m.}}) =
2 m^3 \sqrt{e E_{\mathrm{c.m.}}} 
\int d^3R' d^3r' d^3R d^3r d^{3}R_{A-2}
\nonumber \\ & \times
 \sum_{\alpha} \Big[
\langle \Psi_i | j_L^{\mu\dagger} | \bs{R}', \bs{r}', \bs{R}_{A-2}, \alpha \rangle
\nonumber \\ &\times
g(e,E_{\mathrm{c.m.}},|\bs{r}-\bs{r}'|,|\bs{R}-\bs{R}'|)
\langle \bs{R}, \bs{r} , \bs{R}_{A-2} , \alpha | 
 j_R^\nu | \Psi_i \rangle \Big]\, ,
\end{align}
where $\alpha$ denotes the spin and isospin degrees of freedom of all $A$ nucleons, with a real-time propagator now given by:
\begin{align}
&g(e,E_{\mathrm{c.m.}},|\bs{r}-\bs{r}'|,|\bs{R}-\bs{R}'|)
\nonumber \\ &=
\frac{1}{4\pi^4}
 j_0\left(\sqrt{4m E_{\mathrm{c.m.}}}|\bs{R}'-\bs{R}|\right)
j_0\left(\sqrt{me}\left|\bs{r}'-\bs{r}\right|\right)\, ,
\end{align}
where $j_0$ is the zeroth-order spherical Bessel function.

\section{Relativistic Final-States Kinematics}
\label{sec:relativistic_kinematics}

We now turn our attention to relativistic effects in the energies of the active nucleons in the final state. Since we are restricting our attention to the quasi-elastic peak, the dominant contribution to the cross section arises from electrons scattering from low-momentum nucleons in the ground state. Relativistic effects in the treatment of the ground state wave function are therefore not considered. They may become important in kinematics that probe large-momentum initial-state nucleons, for example in experiments studying short-range correlations~\cite{Hen:2016kwk}. Similarly, we do not include relativistic corrections to the final-state wave functions, but only to the energies of the active nucleons in the final state. 

Kinematical relativistic effects of this kind cannot be easily introduced in most \textit{ab initio} methods, such as the GFMC, which are inherently nonrelativistic. The STA, however, allows their implementation due to the factorization of the final states. Specifically, relativistic expressions for the energies of the two active nucleons are introduced in the energy-conserving delta function $\delta(\omega+E-e-E_{\mathrm{c.m.}})$ appearing in Eq.~\eqref{eq:integration_eE}. This requires expressing the response density in terms of the single-particle momenta (or energies) rather than the corresponding relative and c.m. variables. The latter could be recovered with the inclusion of the angular dependence between the nucleon momenta, a development that is underway.

With the free-propagator, the response function in terms of the single-nucleon momenta, $p_1$ and $p_2$,  becomes: 
\begin{align}
\label{eq:integration_k1k2}
    \begin{split}
    R^{\mu\nu}=&\int_0^\infty dp_1 \int_0^\infty dp_2 D^{\mu\nu}(p_1,p_2)
    \\
    &\times\delta\left(\omega+E-\sqrt{p_1^2+m^2}-\sqrt{p_2^2+m^2}+2m\right)\, ,
    \end{split}
\end{align}
where the response density reads:
\begin{align}
\label{eq:density_p1p2}
D^{\mu\nu}&(p_1,p_2) =
p_1^2 p_2^2
\sum_{\alpha} \int d^3r_1' d^3r_2' d^3r_1 d^3r_2 d^3R_{A-2}
\nonumber \\ & \times
\langle \Psi_i | j_L^{\mu\dagger} | \bs{r}_1', \bs{r}_2', \bs{R}_{A-2}, \alpha \rangle
\nonumber \\ &\times
g(p_1,p_2,|\bs{r}_1-\bs{r}_1'|,|\bs{r}_2-\bs{r}_2'|)
\nonumber \\ &\times
\langle \bs{r}_1, \bs{r}_2 , \bs{R}_{A-2}, \alpha| 
 j_R^\nu | \Psi_i \rangle\, ,
\end{align}
with the free-propagator given by:
\begin{align}
\label{eq:2b_prop_p1p2_nofsi}
&g(p_1,p_2,|\bs{r}_1-\bs{r}_1'|,|\bs{r}_2-\bs{r}_2'|)
\nonumber \\ &=
\frac{1}{4\pi^4}
 j_0(p_1|\bs{r}_1'-\bs{r}_1|)
 j_0(p_2|\bs{r}_2'-\bs{r}_2|)\, .
\end{align}
If nonrelativistic energies were used in Eq.~\eqref{eq:integration_k1k2}, the expression would reduce exactly to Eq.~\eqref{eq:integration_eE}.
The simultaneous inclusion of correlated-propagator with relativistic kinematical effects is deferred to future work.

\begin{figure*}
    \centering
    \includegraphics[width=1\linewidth]{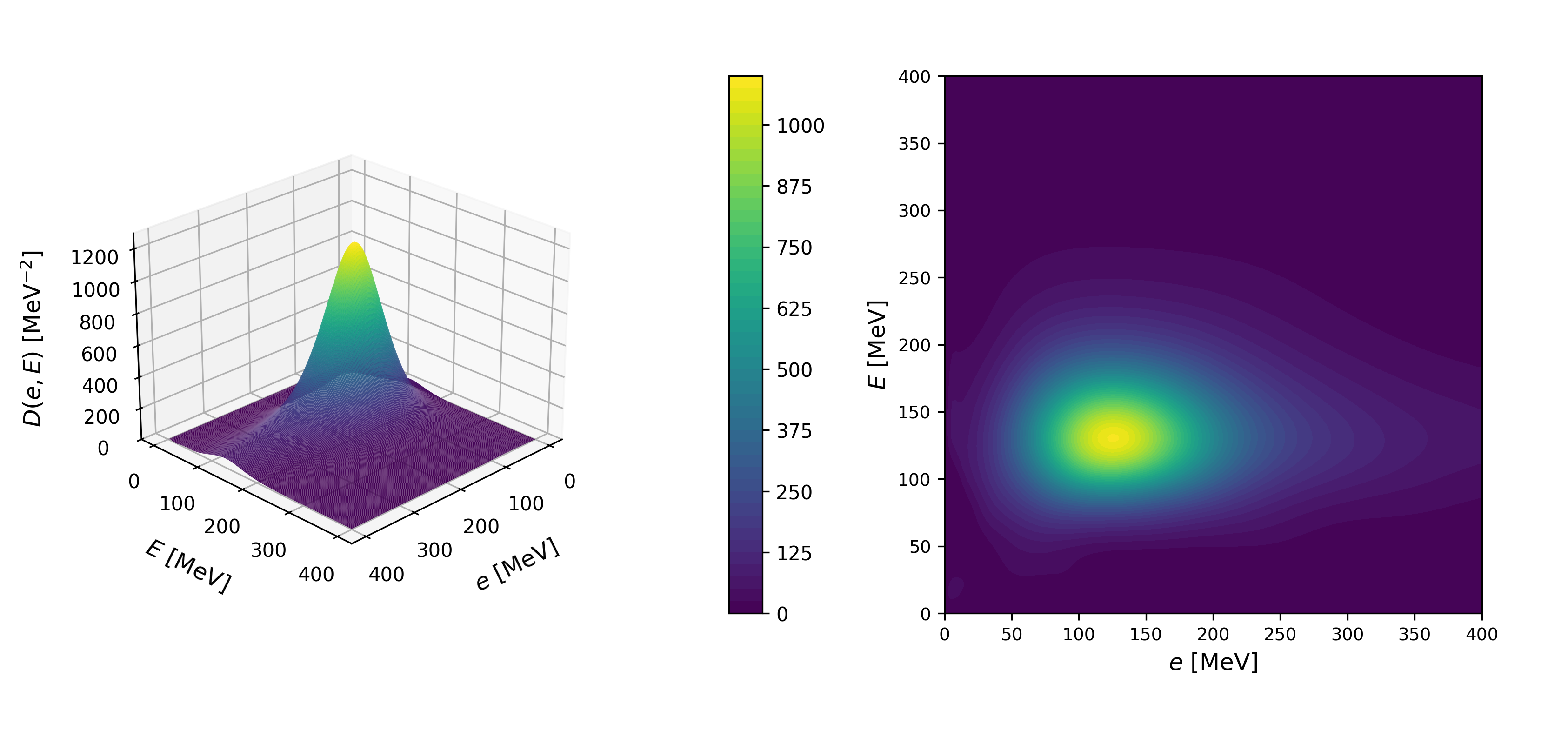}
    \caption{Left panel: Transverse response density for $^4$He at $q=700$ MeV/$c$, as a function of relative and center of mass energies, $e$ and $E$. Results are from Ref.~\cite{Pastore:2019urn} and include the correlated-propagator along with one- and two-nucleon currents. Right panel: Contour plot of the figure in the left panel. See text for explanations.}
    \label{fig:density_eE}
\end{figure*}

\begin{figure*}
    \centering
    \includegraphics[width=1\linewidth]{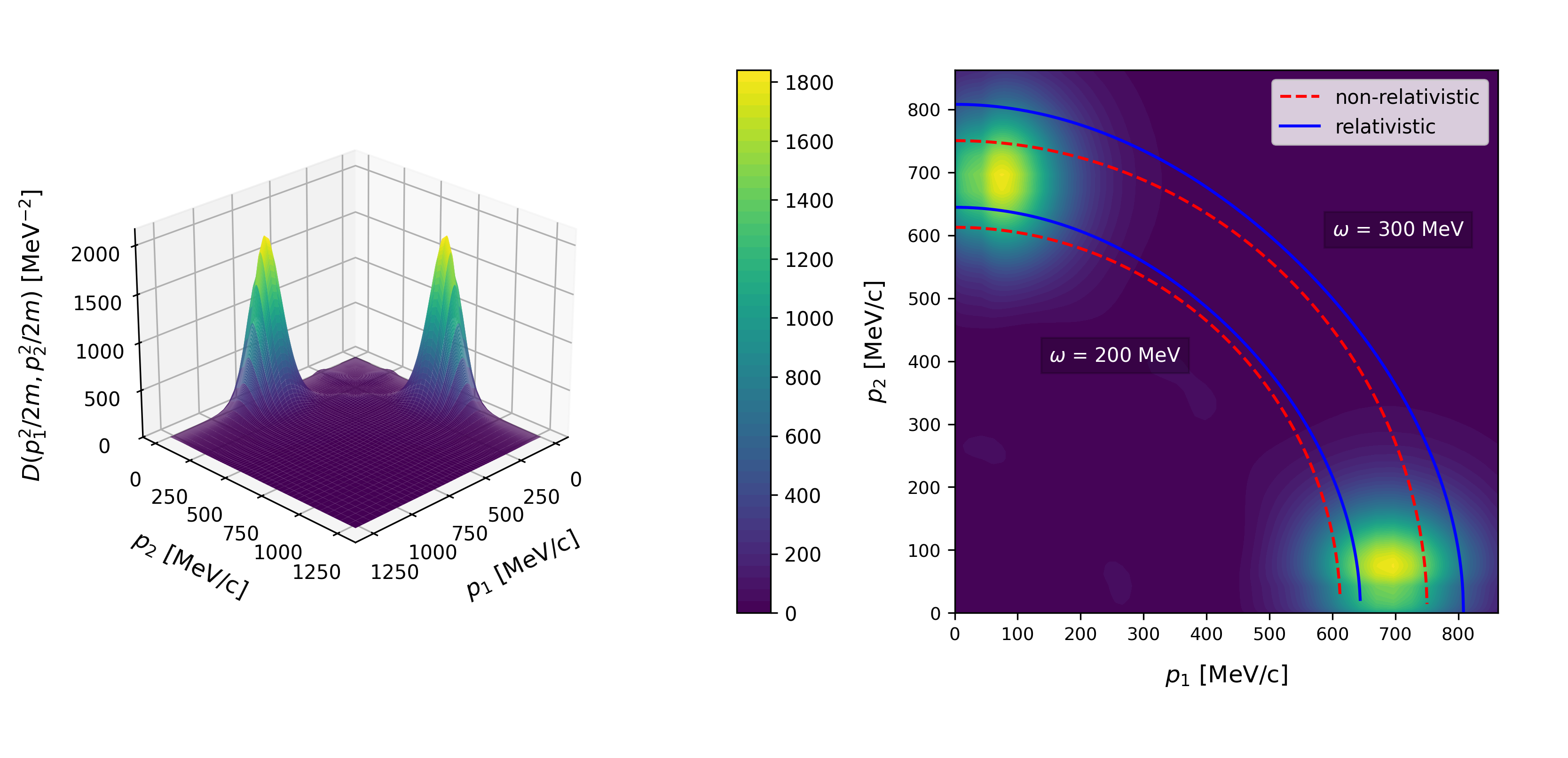}
    \caption{Left panel: Transverse response density for $^4$He at $q=700$ MeV/$c$, as a function of single nucleon momenta, $p_1$ and $p_2$. Results based on the free-propagator and single-nucleon currents up to $(p/m)^1$ in the high-$q$ expansion -- see Eqs.~\eqref{eq:current_p0} and~\eqref{eq:current_p1}.
    Right panel: Contour plot of the figure in the left panel. Solid blue (dashed red) lines indicate the energy-conserving delta function with relativistic (nonrelativistic) energies for the active nucleons in the final state for two given values of energy transfer $\omega$. See text for explanations.}
    \label{fig:density_e1e2}
\end{figure*}

The two response densities, $D^{\mu\nu}(e,E_{\mathrm{c.m.}})$ and $D^{\mu\nu}(p_1,p_2)$, exhibit different behaviors. In the quasi-elastic regime, the dominant contribution to the cross section is due to the interaction of the probe with a single low-momentum nucleon. Consequently, one of the active nucleons carries a momentum $\sim q$ after the interaction, while the other remains unperturbed. As a result, $D^{\mu\nu}(e,E_{\mathrm{c.m.}})$ exhibits a single peak around $e=E=q^2/4m$, while $D^{\mu\nu}(p_1,p_2)$ has two peaks, one around $p_1=q$ and $p_2=0$, and the other around $p_2=q$ and $p_1=0$. These features are displayed in Figs.~\ref{fig:density_eE} and~\ref{fig:density_e1e2} for the transverse response densities in $^4$He at $q=700$ MeV/$c$.

When folded with the appropriate energy-conserving delta function -- either using relativistic or nonrelativistic energies for the two active nucleons in the final state, the integrated response densities give rise to response functions with distinct features. The solid blue and dashed red lines in the right panel of Fig.~\ref{fig:density_e1e2} indicate the positions where the conditions $\omega+E=\sqrt{p_1^2+m^2}+\sqrt{p_2^2+m^2}-2m$ and $\omega+E=p_1^2/2m +p_2^2/2m$, respectively, are satisfied for a given value of $\omega$. In the relativistic case, the peak of the response function is obtained at $\omega\sim\sqrt{q^2+m^2}-m$ (neglecting binding effects), compared with $\omega \sim q^2/2m$ in the nonrelativistic case. Thus, relativistic kinematics shift the peak to lower values $\omega$, with the shift increasing for large values of $q$.

The shape of the peak is also affected. In the nonrelativistic limit, the width of the quasi-elastic peak is given by $\Delta \omega\sim 2\,pq/m$, where $p$ is the typical momentum of the nucleon inside the nucleus, while in the relativistic case the width narrows to $\Delta \omega\sim 2\,pq/\sqrt{q^2+m^2}$.
The integral of the response function over $\omega$ (\textit{i.e.}, the sum rule) remains  unchanged by these relativistic kinematical effects, since the delta function does not alter the result of the integration as long as $D^{\mu\nu}$ does not depend on $\omega$ (which holds in this work based on $\omega$-independent currents and nucleonic form factors). As a result, the narrowing of the peak implies a corresponding increase of its height to conserve the sum rule. In summary, relativistic kinematics effects for final-state energies lead to a shift toward lower values of $\omega$, with increased peak height and reduced width.

\section{Results}
\label{sec:results}
In this section, we present our results for both response functions and inclusive double differential cross sections for electron scattering from $^3$He and $^4$He based on the \textit{high-$q$ nonrelativistic reduction of the single-nucleon covariant electromagnetic current} and \textit{inclusive of relativistic kinematics in the two-nucleon final states}. These results will be denoted in what follows with `STA($n$)' with $n=0$ or $1$ depending on whether we use the one-body currents up to $(p/m)^0$ -- Eqs.~\eqref{eq:rho_p0} and~\eqref{eq:current_p0} -- or $(p/m)^1$ -- Eqs.~\eqref{eq:rho_p1} and~\eqref{eq:current_p1} -- corrections in the high-$q$ expansion scheme. The STA($n$) results are obtained using the free-propagator implying an inconsistency in the implementation of the interference term, discussed in detail in Ref.~\cite{Pastore:2019urn}. In order to isolate the effect of using relativistic final-state kinematics, we will denote with `STA($n$)NR$\delta$' results that use nonrelativistic kinematics in the delta conserving delta function, corresponding to kinematics indicated by the red dashed line in the right panel of Fig.~\ref{fig:density_e1e2}. 

We will compare the STA($n$) with those of Refs.~\cite{Pastore:2019urn, Andreoli:2021cxo} obtained using the correlated-propagator along with one- and two-body electromagnetic currents, denoted with `STA'. We stress that the STA calculations are based on the low-$q$ nonrelativistic reduction of the single-nucleon current at LO, Eqs.~\eqref{eq:charge_NR_LO} and~\eqref{eq:transverse_NR_LO}, with the charge operator including also the Darwin-Foldy and spin-orbit terms in Eq.~\eqref{eq:charge_DF}.

\begin{widetext}
    \begin{center}
    \includegraphics[width=0.48\textwidth]{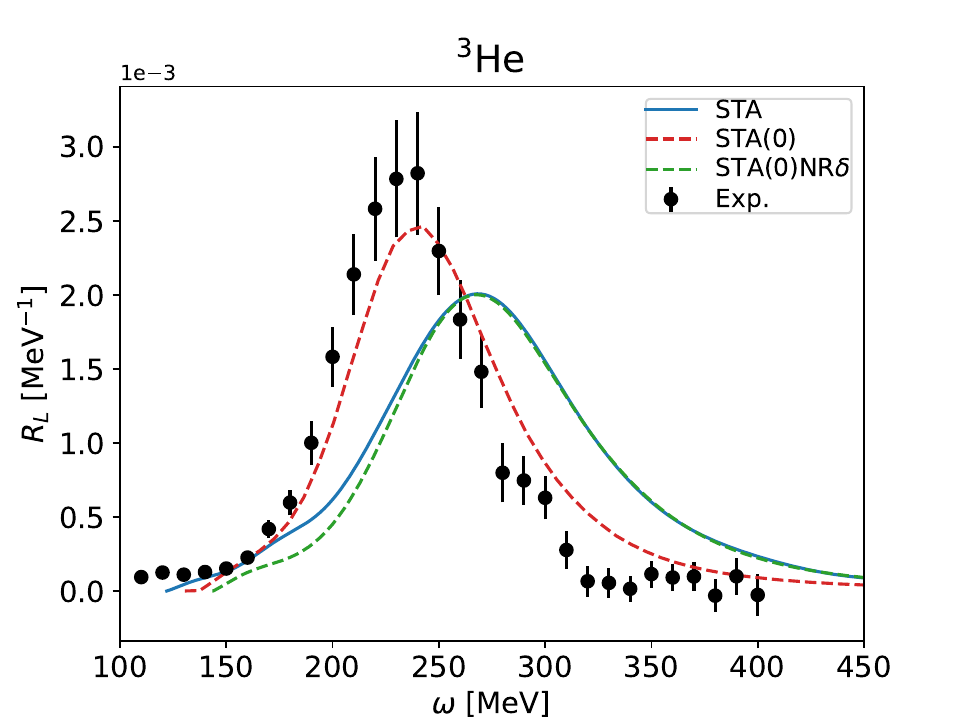}
    \includegraphics[width=0.48\textwidth]{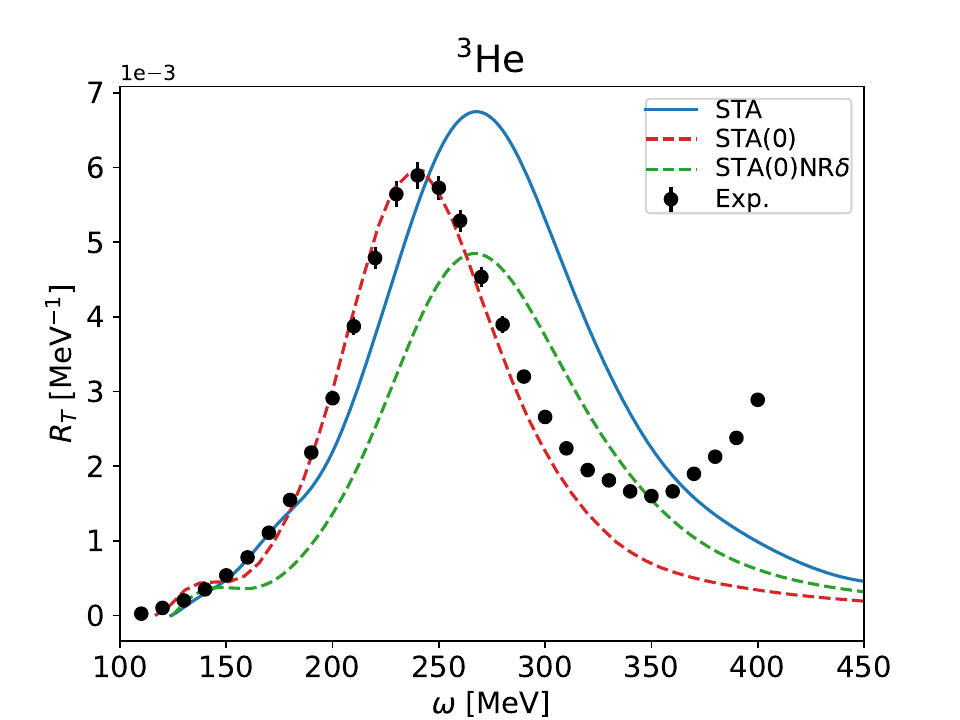}
        \captionof{figure}{Longitudinal (left) and transverse (right) response functions for $^3$He at $q=700$ MeV/$c$. Solid blue and dashed red lines show STA and STA(0) results, respectively. Green dashed lines correspond to STA(0)NR$\delta$ results.   Experimental data from Ref.~\cite{Carlson:2001mp} are given by the black symbols.}
    \label{fig:responses_delta}
\end{center}
\end{widetext}

\subsection{High-$q$ expansion scheme and relativistic final-state kinematics}

In order to analyze the combined impact of the high-$q$ expansion scheme and the relativistic final-state kinematics, we examine the $^3$He response functions at $q=700$ MeV/$c$, where the conventional STA failed to explain the data~\cite{Andreoli:2021cxo}. We begin with the longitudinal response function shown in the left panel of Fig.~\ref{fig:responses_delta} along with the experimental data. The conventional STA calculation is displayed by the solid blue line. To isolate the effect of the high-$q$ expansion scheme for the single-nucleon current operator, we can compare this with the STA(0)NR$\delta$ calculation (green dashed line), which utilizes the same nonrelativistic final state kinematics as the original STA. Thus, the only difference lies in the implementation of the one-body charge operator. The STA uses the low-$q$ expansion scheme at LO plus the Darwin-Foldy and spin-orbit terms, whereas the STA(0)NR$\delta$  retains $(p/m)^0$ term in the high-$q$ expansion scheme. As discussed in Sec.~\ref{sec:comparing} and illustrated in Fig.~\ref{fig:R_L_factors}, the difference between the two formulations is negligible at $q=700$ MeV/$c$, as in this region the Darwin-Foldy term and the leading $(p/m)^0$ charge operator yield the same behavior. Substantial differences are expected at higher values of momentum transfer. To highlight the role of the relativistic final-state kinematics, we next compare the STA(0)NR$\delta$ calculation with the STA(0) result (red dashed line), which incorporates relativistic kinematics in the energy-conserving delta function. This leads to a shift of the strength to lower values of $\omega$, along with a narrowing of the width and an enhancement of the peak's height, resulting in an improved description of the position of the peak. 

The transverse response function is shown in the right panel of Fig.~\ref{fig:responses_delta}, with the same notation as in the left panel. In this case, the suppression of the STA(0)NR$\delta$ calculation with respect to the original STA result is more pronounced. As shown in Fig.~\ref{fig:R_T_factors}, at $q=700$ MeV/$c$, the STA(0)NR$\delta$ is suppressed by a factor $\sim 0.7$ with respect to the original STA. Also in this case, inclusion of relativistic final-state kinematics leads to a very nice agreement with the experimental data, a feature not displayed by the original STA calculation. The latter is limited by the assumption that $q \ll m$, and we have verified that the inclusion of the  sub-leading $(q/m)^2$ terms cannot overcome this intrinsic limitation. This case highlights the need for both the high-$q$ expansion scheme and the relativistic treatment of final-state kinematics.

\subsection{Convergence of the high-$q$ expansion scheme}

\begin{figure*}
    \centering
    \includegraphics[width=0.49\linewidth]{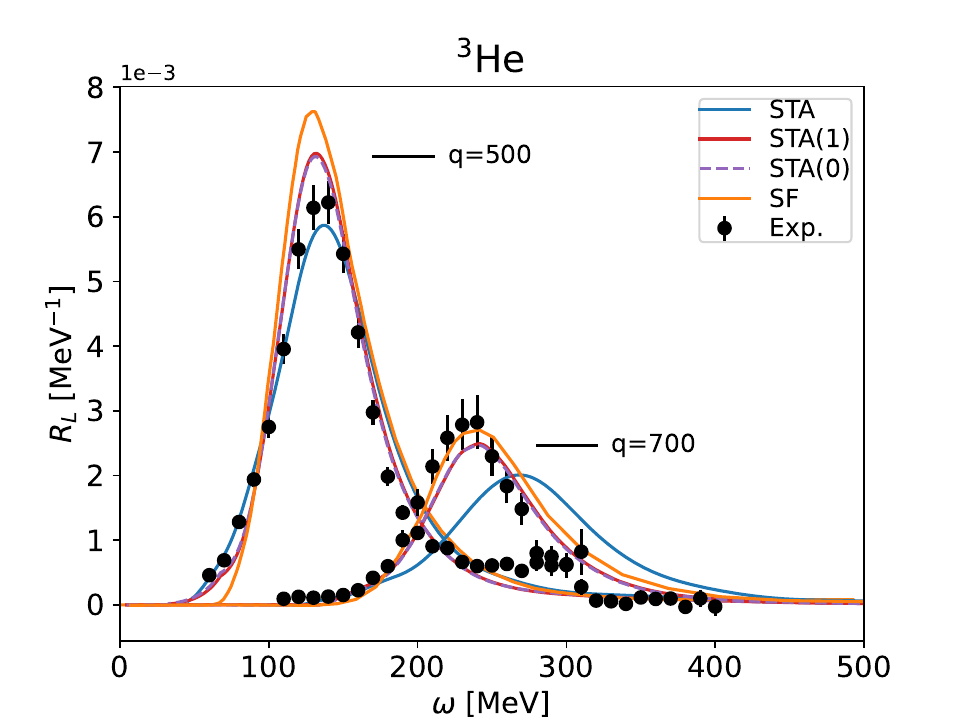}
    \includegraphics[width=0.49\linewidth]{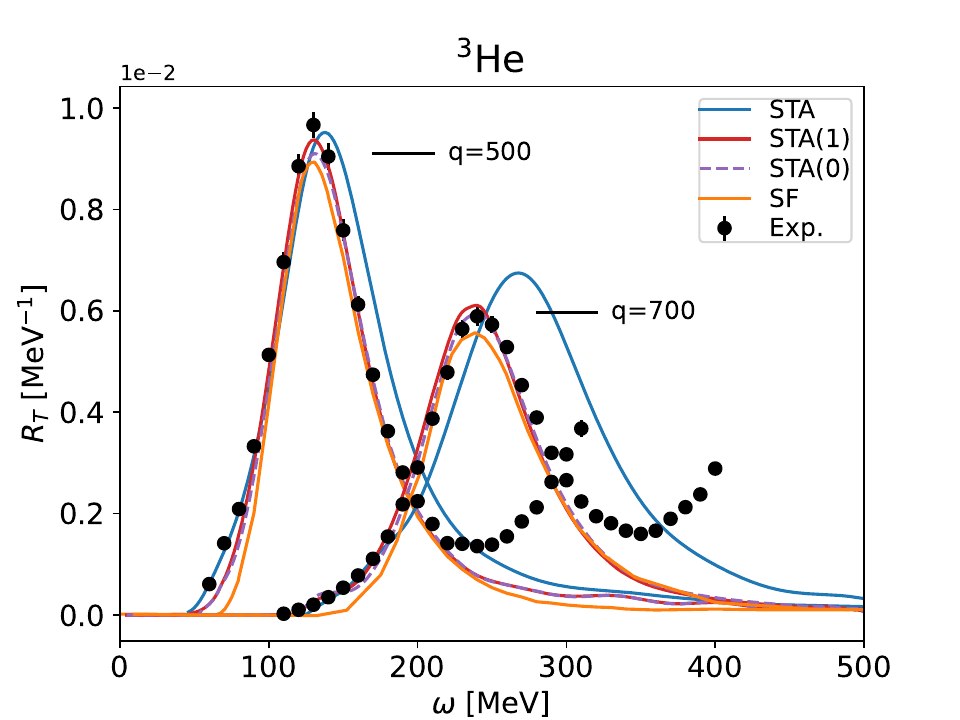}
    \centering
    \includegraphics[width=0.49\linewidth]{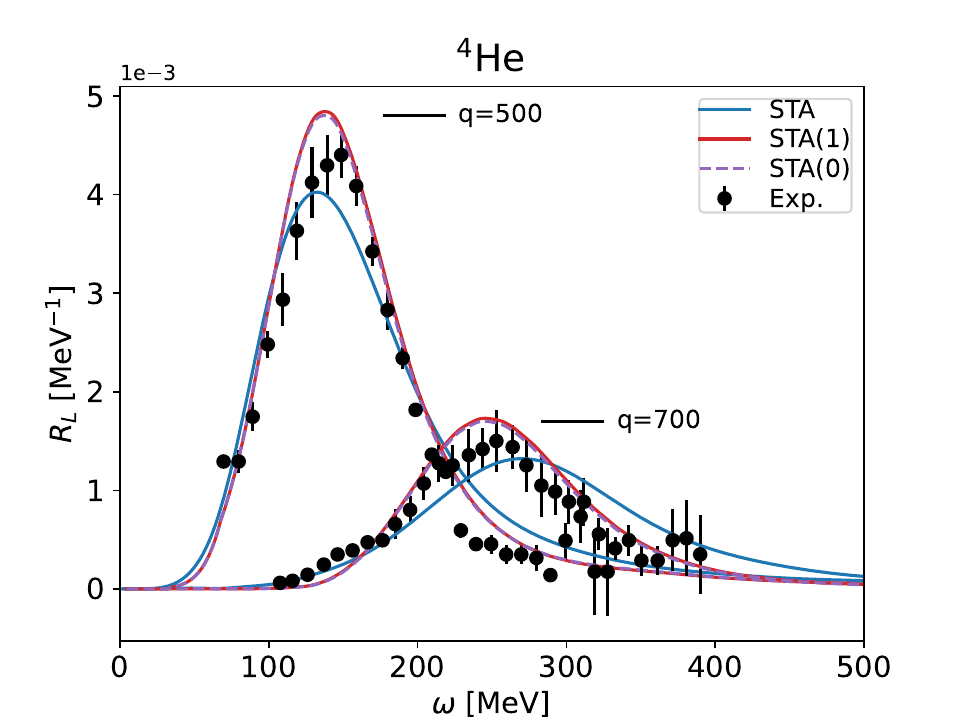}
    \includegraphics[width=0.49\linewidth]{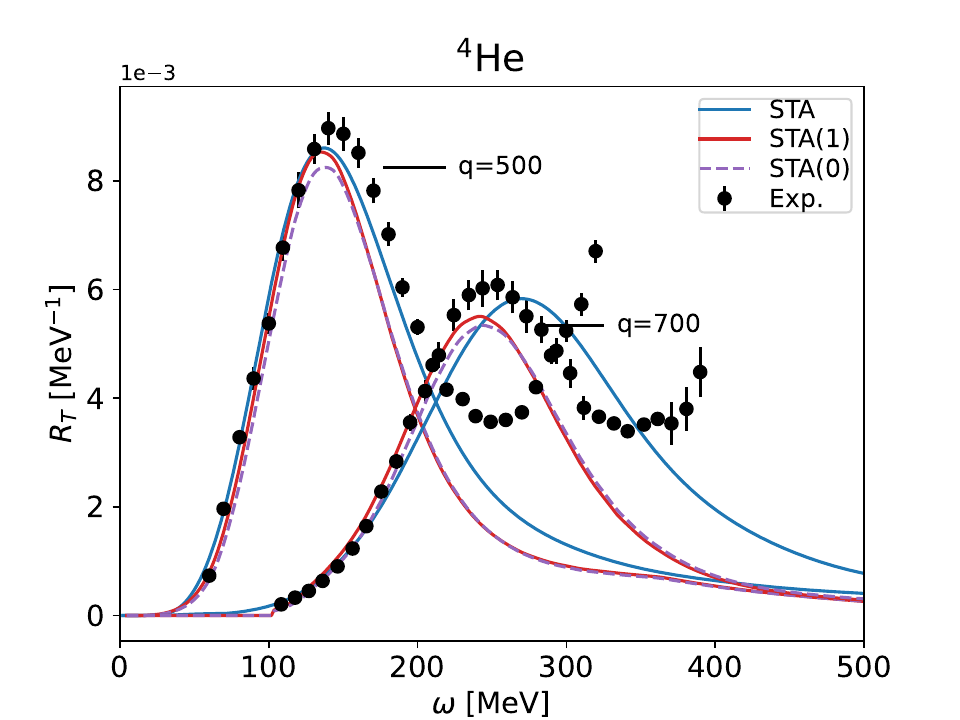}
    \caption{Longitudinal (left) and transverse (right) response functions for $^3$He (top panels) and $^4$He (bottom panels), at $q=500$ and $q=700$ MeV/$c$. The solid blue lines are STA calculations without the relativistic treatment discussed in this work. The solid red (dashed purple) lines use the new currents up to NLO (LO). Orange solid lines show the calculations of the spectral function approach for $^3$He from Ref.~\cite{Andreoli:2021cxo}. Experimental data is from Ref.~\cite{Carlson:2001mp}.}    \label{fig:responses_final}
\end{figure*}

Next, we examine the convergence of the high-$q$ expansion scheme by comparing the STA(0) and STA(1) calculations for the $^3$He and $^4$He response functions at two values of momentum transfer, namely $q=500$ and $700$ MeV/$c$. The $^3$He ($^4$He) longitudinal and transverse response functions are shown in the top (bottom) panels of Fig.~\ref{fig:responses_final}, along with the experimental data. In the case of $^3$He, we also show the Spectral Function results from Ref.~\cite{Andreoli:2021cxo} (orange solid lines). 

The original STA calculations are given by the blue solid lines, while STA(0) and STA(1) calculations are shown by the purple dashed lines and red solid lines, respectively. In all the considered cases, there is a negligible difference between the STA(0) and STA(1) calculations, implying that the high-$q$ expansion is rapidly converging. 

The difference between the STA and STA($n$) calculations is clearly more remarked at $q=700$ MeV/$c$ where relativistic effects are more pronounced. Here is where relativistic corrections, both in the single-nucleon currents and final-state energies, are required to reach an improved agreement with the data. 

The STA(1) under-prediction of the $^4$He transverse response function at $q=700$ MeV/$c$ is due to pion-production mechanisms currently lacking in our formalism. At $q=500$ MeV/$c$, both the $^3$He and $^4$He longitudinal response functions are slightly over-predicted by the STA(1) formulation. This is likely due to the use of the free-propagator as opposed to the correlated-propagator. Indeed, at $q=500$ MeV/$c$ where relativistic effects are less pronounced, the original STA calculations -- based on the correlated-propagator -- are slightly suppressed with respect to the STA(1) results.

\subsection{Cross Sections}
To calculate double differential cross sections, response functions are evaluated for a set of values of $q$ in the range 300-1400 MeV/$c$, using the interpolation scheme described in Ref.~\cite{Andreoli:2024ovl}. This allows us to generate response functions for arbitrary values of $q$ on a finer grid required for the determination of the cross section.  In Figs.~\ref{fig:cross_sections_he3_p1} and~\ref{fig:cross_sections_he4_p1}, we present cross sections for electron scattering from $^3$He and $^4$He, for selected values of beam energies and scattering angle. Solid red lines correspond to STA(1), while dashed blue lines correspond to the original STA calculations. The figures also show the position of the peak with a vertical black dashed line. 

In all the considered kinematics, the inclusion of relativistic final-state energies improves the description of the position of the quasi-elastic peak. Their main effect is a redistribution of the cross section strength towards the `left' and closer to the observed peak. These effects are clearly more pronounced as the position of the peak corresponds to higher values of $q$.  In backward scattering ({\it e.g.}, $\theta=144.5^\circ$), where the cross section is dominated by the transverse response function, the STA(1) results are suppressed with respect to the original calculations due to the suppression factor induced by the high-$q$ expansion scheme adopted for the single-nucleon current. 

Overall, STA(1) calculations show good agreement with the experimental data in the quasi-elastic region, for all the kinematics considered. However, they do not reproduce experimental data at higher values of $\omega$ as the current formalism does not include resonances and meson-production. 
Compared to the original STA calculations, the present STA(1) results provide a more accurate description of the experimental data, as the apparent better agreement of previous STA calculations for certain cross sections (see {\it e.g.}, results in Fig.~\ref{fig:cross_sections_he4_p1} for $\epsilon=0.5-0.6$ GeV) 
 would be compromised by the inclusion of pion production mechanisms.

In some instances, the STA(1) results slightly over-predict the height of the peak. This behavior requires further attention, but it is likely attributable to the fact that the STA(1) results are based on the free-propagator and are missing the effect of the two-nucleon correlations at the vertex. The inclusion of final-state two-nucleon interactions is expected to reduce the height of the peak and improve agreement with experiment. Additionally, the current treatment lacks consistent relativistic one- and two-body currents, as we employ relativistic one-body currents while retaining nonrelativistic two-body currents from previous calculations. Work along these lines is on the way. 

\begin{figure*}
    \centering
    \includegraphics[width=1\linewidth]{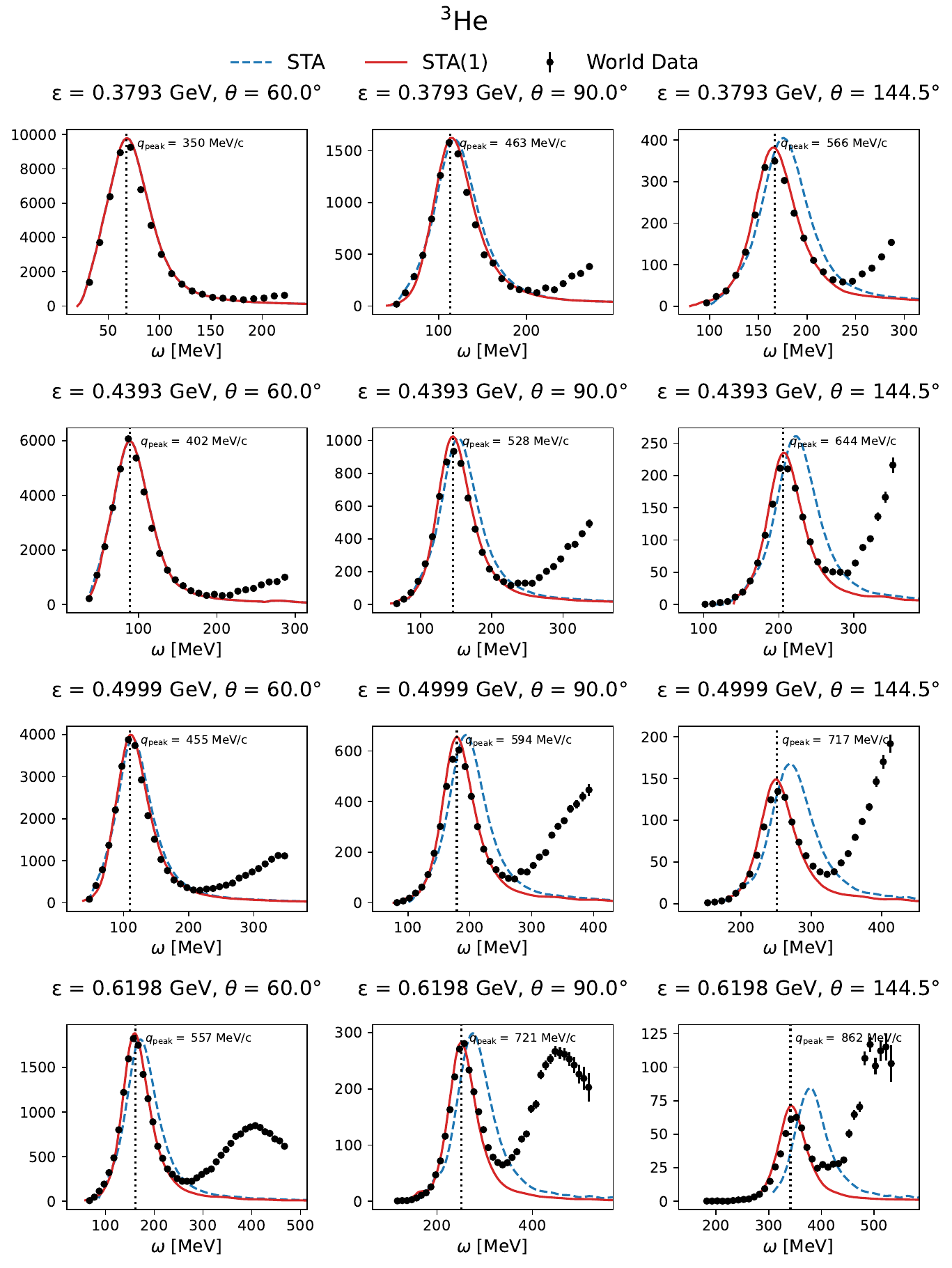}
    \phantomcaption
\end{figure*}
\begin{figure*}\ContinuedFloat
    \centering
    \includegraphics[width=1\linewidth]{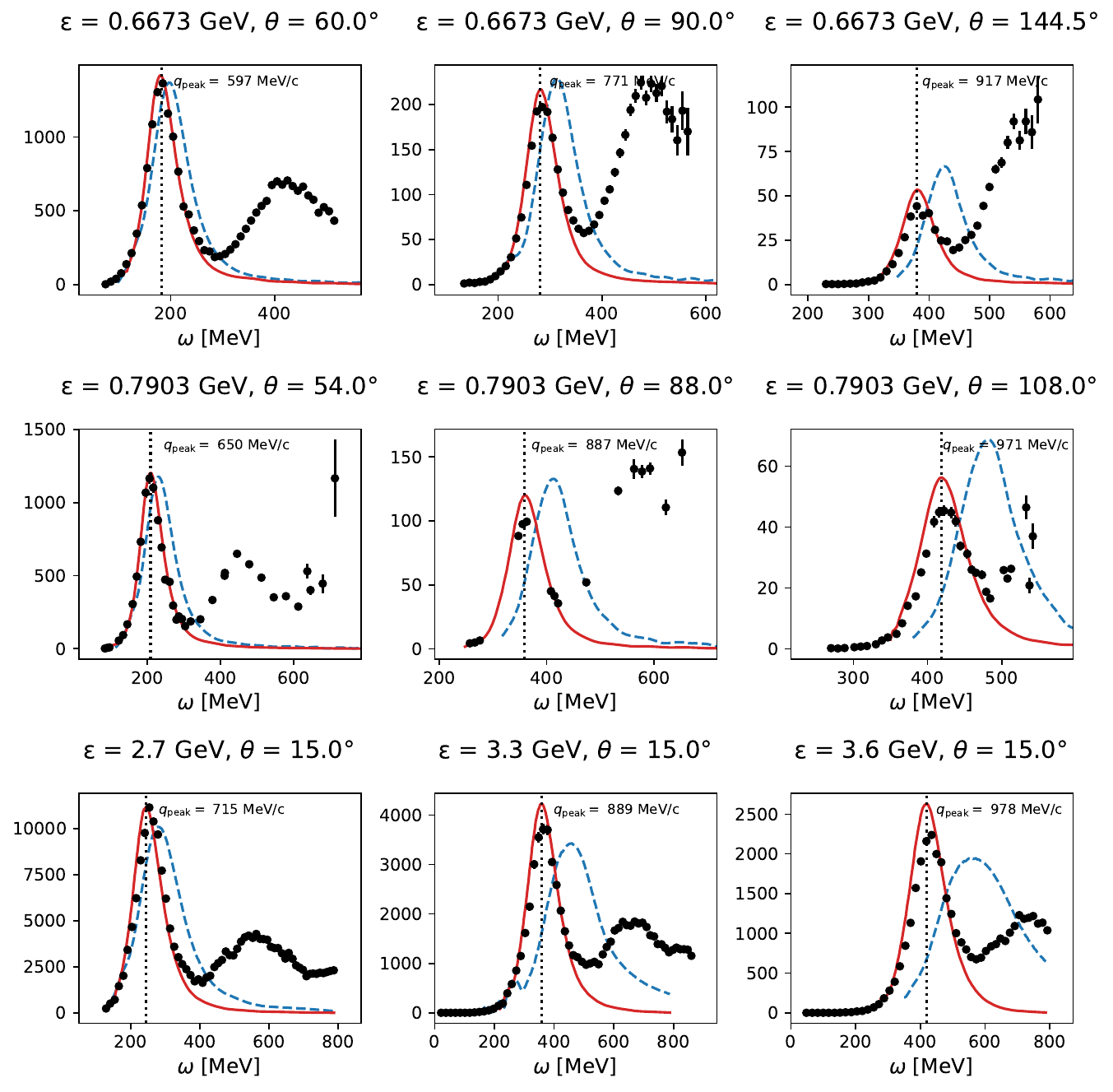}
    \caption{Inclusive double-differential cross sections for electron scattering on $^3$He for different values of beam energy $\varepsilon$ and scattering angle $\theta$ . The dashed blue (solid red) lines are STA calculations without (with) the relativistic treatment discussed in this work. Units are in nb/sr GeV. Experimental data is from~\cite{virginiaarchive}.}    \label{fig:cross_sections_he3_p1}
\end{figure*}
\begin{figure*}
    \centering
    \includegraphics[width=1\linewidth]{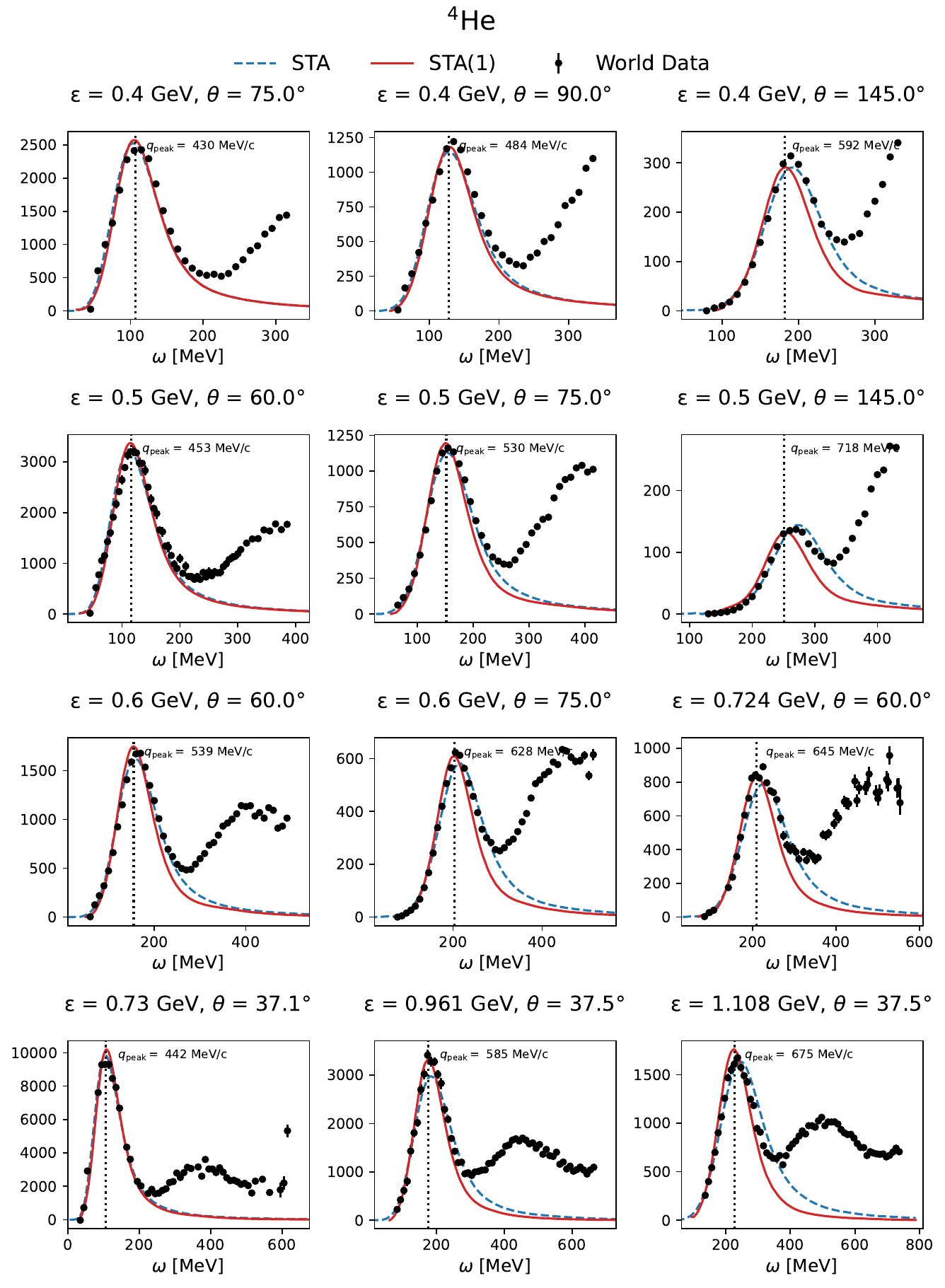}
    \caption{The same as Fig. \ref{fig:cross_sections_he3_p1}, but for $^4$He.}
    \label{fig:cross_sections_he4_p1}
\end{figure*}

\section{Conclusions}
\label{sec:conclusion}
In this work, we have presented an extension of the STA that includes relativistic effects and allows us to provide accurate predictions for large-momentum-transfer quasielastic reactions. Such effects enter via the currents, describing the interaction of the lepton with the nucleus, as well as via the final-state energy of struck nucleons.

For the currents, an expansion of the covariant one-body expression is used, in which only the initial momentum of the nucleon is assumed to be small. It provides an accurate description for quasielastic kinematics for any value of momentum transfer $q$, unlike the more traditional expansion that is valid only for small values of $q$. We have shown that it results in a significant reduction of the calculated response functions, compared to the traditional nonrelativistic expansion. This high-$q$ expansion can be used in various applications involving electromagnetic currents, not only with the STA and not only for lepton-nucleus cross sections. Similar expansion can be derived for electroweak currents.

The STA, due to the assumed factorization of the final states, allows us to also account for the kinematical relativistic effects of the struck nucleons. Working with single-particle momenta and neglecting the impact of FSI, we have reformulated the appropriate equations of the STA to incorporate relativistic expressions for the kinetic energy of the active pair in the final state. We have discussed its impact on the location, width, and amplitude of the quasielastic peak. We plan to combine the impact of FSI and kinematical relativistic effects in future work, although the impact of FSI seems to be small for large enough values of $q$.

We have performed full STA calculations of inclusive electron scattering with these relativistic effects for $^3$He and $^4$He. First, we have demonstrated the separate impact of the two effects. We have seen that the use of the new expressions for the currents leads to the expected effect on the amplitude of the response, specifically, a significant reduction for the transverse response for $q=700$ MeV/c; a significant effect is also observed for the longitudinal response for larger values of $q$, compared to the Darwin-Foldy term. Then, the inclusion of relativistic expressions for the final-state energies of the pair shifts the location of the peak and increases its magnitude, leading to an agreement with experimental data.

We also studied the convergence of the high-$q$ expansion of the currents, demonstrating that the impact of NLO terms is small. 
Finally, we have presented full calculations for response functions as well as inclusive cross sections for both $^3$He and $^4$He. The missing impact of FSI is seen in the longitudinal responses at $q=500$ MeV/c, and the missing impact of pion production is seen in the transverse response (particularly for $^4$He at $q=700$ MeV/c). Otherwise, good agreement with the data is observed, with significant improvements compared to calculations that do not incorporate relativistic effects. 

These advances are important to support experimental programs in neutrino oscillations as well as nuclear structure studies via large-momentum-transfer quasielastic reactions, as used, for example, in the study of short-range correlations.

\acknowledgments
We are thankful for useful discussions with S. Jeschonnek.
We also thank G.~B.~King, M.~Macedo-Lima, J.~Carlson and R.~B.~Wiringa for interesting discussions at various stages of this work.
L.~A. and R.~W. contributed equally to this work.
This work is supported by the US Department of Energy under Contracts No. DE-SC0021027 (G.~C.-W., S.~P.), DE-AC05-06OR23177 (L.~A, A.~G), a 2021 Early Career Award number DE-SC0022002 (M.~P.), and the NUCLEI SciDAC program (S.~P., M.~P.). G.~C.-W. acknowledges support from the NSF Graduate Research Fellowship Program under Grant No. DGE-213989. We thank the Nuclear Theory for New Physics Topical Collaboration, supported by the U.S.~Department of Energy under contract DE-SC0023663, for fostering dynamic collaborations. A.~G. acknowledges the direct support of the Nuclear Theory for New Physics Topical collaboration. R.~W. acknowledges support by the Edwin Thompson Jaynes Postdoctoral Fellowship of the Washington University Physics Department. The work of S.G. is supported by the U.S. Department of Energy through the Los Alamos National Laboratory. Los Alamos National Laboratory is operated by Triad National Security, LLC, for the National Nuclear Security Administration of U.S. Department of Energy (Contract No. 89233218CNA000001), by the Office of Advanced Scientific Computing Research, Scientific Discovery through Advanced Computing (SciDAC) NUCLEI program, and by the LANL LDRD program.

The many-body calculations were performed on the parallel computers of the Laboratory Computing Resource Center, Argonne National Laboratory, the computers of the Argonne Leadership Computing Facility (ALCF) via the INCITE grant ``Ab-initio nuclear structure and nuclear reactions'', the 2019/2020 ALCC grant ``Low Energy Neutrino-Nucleus interactions'' for the project NNInteractions, the 2020/2021 ALCC grant ``Chiral Nuclear Interactions from Nuclei to Nucleonic Matter'' for the project ChiralNuc, the 2021/2022 ALCC grant ``Quantum Monte Carlo Calculations of Nuclei up to $^{16}{\rm O}$ and Neutron Matter'' for the project \mbox{QMCNuc}, and by the National Energy Research
Scientific Computing Center, a DOE Office of Science User Facility
supported by the Office of Science of the U.S. Department of Energy
under Contract No. DE-AC02-05CH11231 using NERSC award
NP-ERCAP0027147 and  NP-ERCAP0031771.

\appendix
\section{Comparison with the Rosenbluth cross section}
\label{sec:ros_cross_section}
In the limit of a single static nucleon in the initial state, our LO expressions for the charge and current operators, Eqs. \eqref{eq:rho_p0} and \eqref{eq:current_p0}, should lead to the well-known Rosenbluth cross section.
In this limit, using Eq. \eqref{eq:rho_p0}, the longitudinal response function becomes 
\be
R_L(q,\omega) = \alpha^2(q) G_E^2 \delta(m+\omega-\sqrt{q^2+m^2}).
\ee
Notice that the particle has momentum $\bs{q}$ in the final state.
The transverse response function becomes
\be
R_T = \left(\frac{2m\tau_{qe}}{q^2}\right)^2 G_M^2 \alpha^2(q) 2q^2 \delta(m+\omega-\sqrt{q^2+m^2})
\ee
using Eq. \eqref{eq:current_p0}. The operator $\bs{\sigma} \times \bs{q} = q (\bs{\sigma} \times \hat{z})$ gives the factor $q^2$. The $\bs{\sigma}$ operator flips the spin but the relevant spin matrix element with the appropriate final state is simply $1$. The factor of $2$ comes from the sum of the $xx$ and $yy$ contributions in Eq. \eqref{eq:R_T}.

Next, the response functions enter Eq. \eqref{eq:diff_cross_section} for the expression of the double differential cross section, and we obtain
\begin{align}
\frac{d^2\sigma}{d\varepsilon'd\Omega'} &= \left(\frac{d\sigma}{d\Omega'}\right)_M  \left[\frac{G_E^2+(1+2(1+\tau_{qe})\tan^2\frac{\theta}{2})\tau_{qe} G_M^2}{(1+\tau_{qe})} \right] 
\nonumber \\ & \times
\frac{1}{2\tau_{qe}+1} 
\delta(m+\omega-\sqrt{q^2+m^2}).
\end{align}
We used here the relations
\be
q^2=4m^2\tau_{qe}(1+\tau_{qe}),
\ee
\be
\frac{Q^2}{q^2}=\frac{1}{1+\tau_{qe}},
\ee
and
\be
\alpha^2(q) =  \frac{\tau_{qe}+1}{2\tau_{qe}+1}.
\ee

To compare with the Rosenbluth cross section, we need to integrate over the electron final energy $\varepsilon'$ to obtain $d\sigma/d\Omega'$. To do that, we need to write the expression that appears inside the delta function using the final and initial energies of the electron, $\varepsilon'$ and $\varepsilon$, respectively. We use
$\omega = \varepsilon - \varepsilon'$ and $q^2 =  \varepsilon^2+\varepsilon'^2-2\varepsilon\varepsilon' cos(\theta)$ (neglecting the mass of the electron), and obtain, after some algebra,
\be
\frac{\partial \varepsilon_1}{\partial \varepsilon'} =
\frac{ \varepsilon}{ \varepsilon'(2\tau_{qe}+1)},\ee
where $\varepsilon_1=\sqrt{m^2+q^2} -\omega$, and the derivative was evaluated at the QE peak (i.e., constrained by the above delta function).
Using this result to change variables in the delta function, the integral over $\varepsilon'$ becomes trivial and we obtain 
\begin{align}
\frac{d\sigma}{d\Omega'} &= \left(\frac{d\sigma}{d\Omega'}\right)_M \frac{\varepsilon'}{\varepsilon} \frac{G_E^2+\left(1+2(1+\tau_{qe})\tan^2\frac{\theta}{2}\right)\tau_{qe} G_M^2}{1+\tau_{qe}}. 
\end{align}
This is exactly the Rosenbluth cross section, which shows that, indeed, in the limit of a static single nucleon in the initial state, our LO expressions reproduce the exact relativistic cross section. This is also a good sanity check for the convention used in the paper for the spinor normalization (Eqs. \eqref{eq:N_p} and \eqref{eq:N_p'}) together with Eq. \eqref{eq:diff_cross_section} for the cross section.

\bibliography{biblio}

\end{document}